\documentclass{emulateapj}

\usepackage{graphics}

\newcommand{\figMagHist}[3]%
{
\begin{#1}[#2]
\begin{centering}
  \label{fig:magDist}
  \includegraphics[width=#3\textwidth]{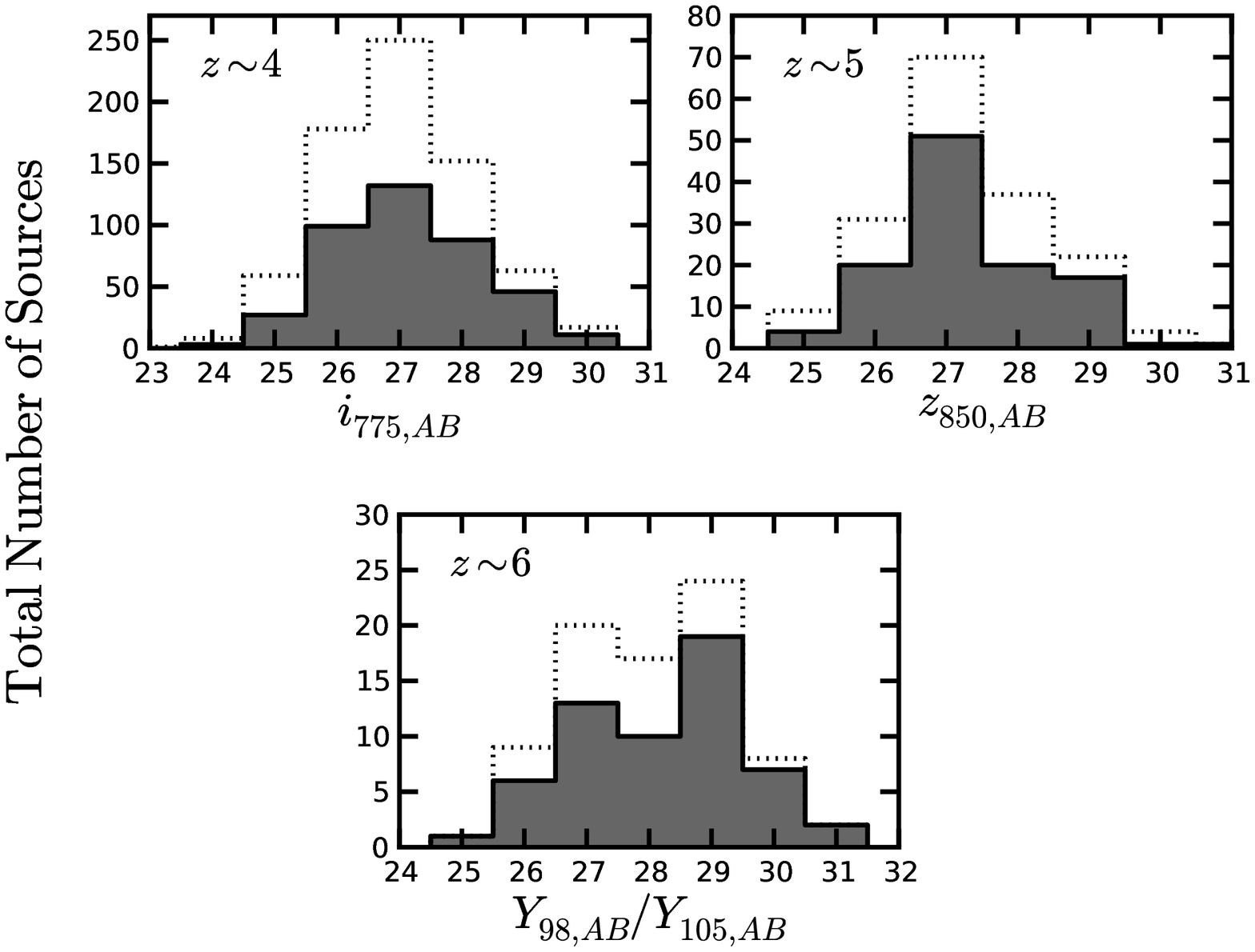}
  \caption{The observed magnitude distribution of the sample.  The
  reference filters for each redshift sub-sample have been chosen to
  be close to rest-frame $\lambda=1500\,$\AA. For the $z\sim6$ sources
  in the ERS we used the $Y_{98}$ filter whereas the $Y_{105}$ filter
  was used for the sources in the HUDF.  The open histograms include
  all the sources and the gray filled histograms include only the
  sources with clean IRAC photometry. This selection does not
  introduce biases relevant to the determination of the mean
  rest-frame optical colors of these sources.}
\end{centering}
\end{#1}
}

\newcommand{\Bstamps}[3]%
{
\begin{#1}[#2]
  \includegraphics[width=#3\textwidth]{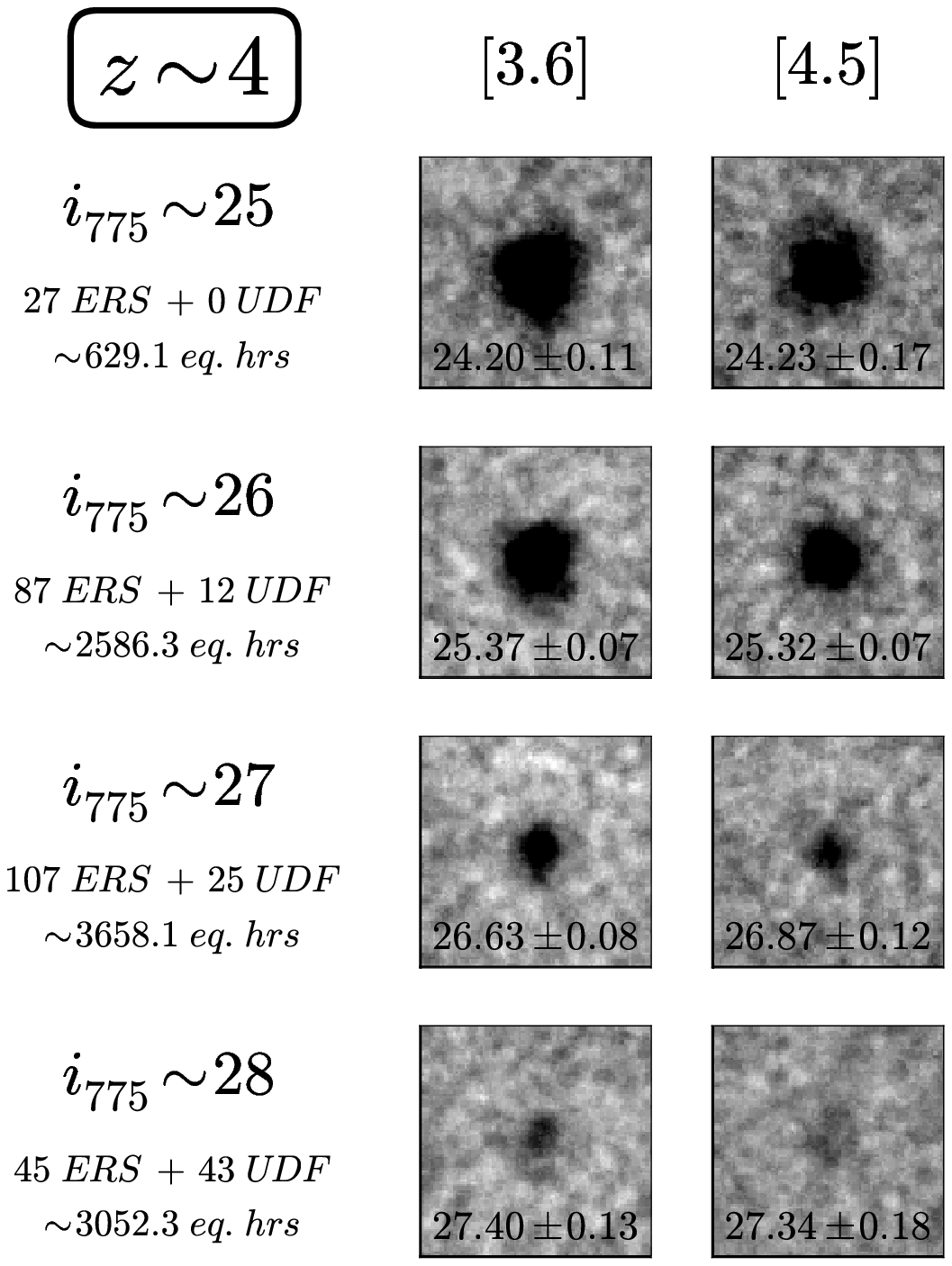}
  \caption{Median stacked images of the $z\sim4$ sources.  The sample
  has been split according to their $i_{775,AB}$ magnitude (approx.
  rest frame 1500\,\AA) in bins of 1 mag. The stamps are 10\arcsec on
  a side.  The number of sources stacked along with the total
  equivalent integration time has been included to the left of the
  stamps.  The magnitudes included in the stamps were measured using
  circular apertures of 2.5\arcsec in diameter and corrected to total
  assuming stellar profiles which amounts to $\sim-0.6$ mags in both
  the [3.6] and the [4.5] channels.  The background was estimated from
  the stacks and subtracted off the images. To estimate the
  uncertainties on the fluxes measured we created 200 random
  realizations of the stack.  In each realization, sources were drawn
  with replacement and then median stacked pixel by pixel.}
  \label{fig:BstackStamp}
\end{#1}
}

\newcommand{\Vstamps}[3]%
{
\begin{#1}[#2]
  \includegraphics[width=#3\textwidth]{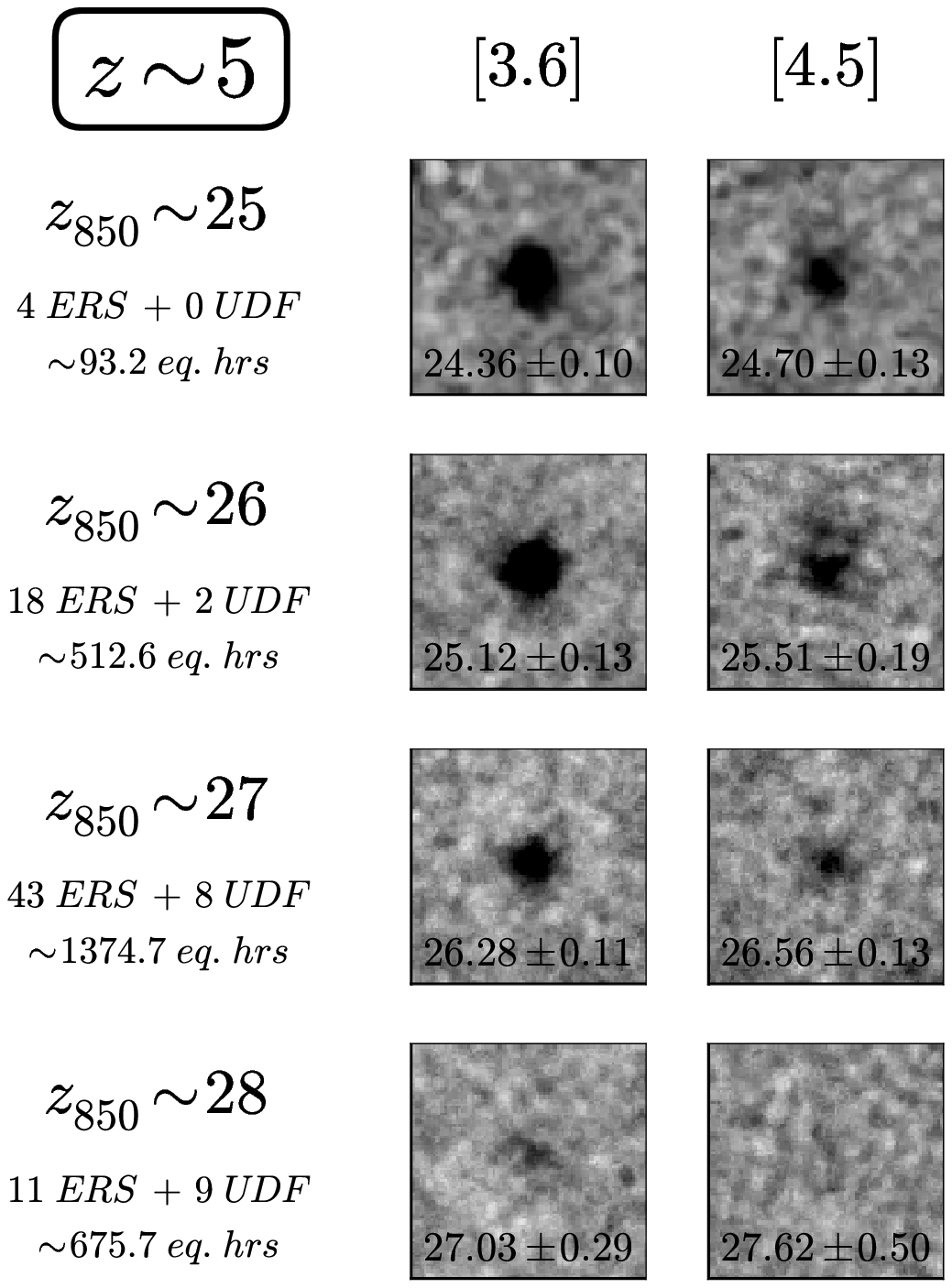}  
  \caption{As in Figure \ref{fig:BstackStamp} but for the $z\sim5$
    sample.  In this case the sample was split according to their
    $z_{850}$ magnitude to approximate rest-frame 1500\,\AA.}
  \label{fig:VstackStamp}
\end{#1}
}

\newcommand{\istamps}[3]%
{
\begin{#1}[#2]
  \includegraphics[width=#3\textwidth]{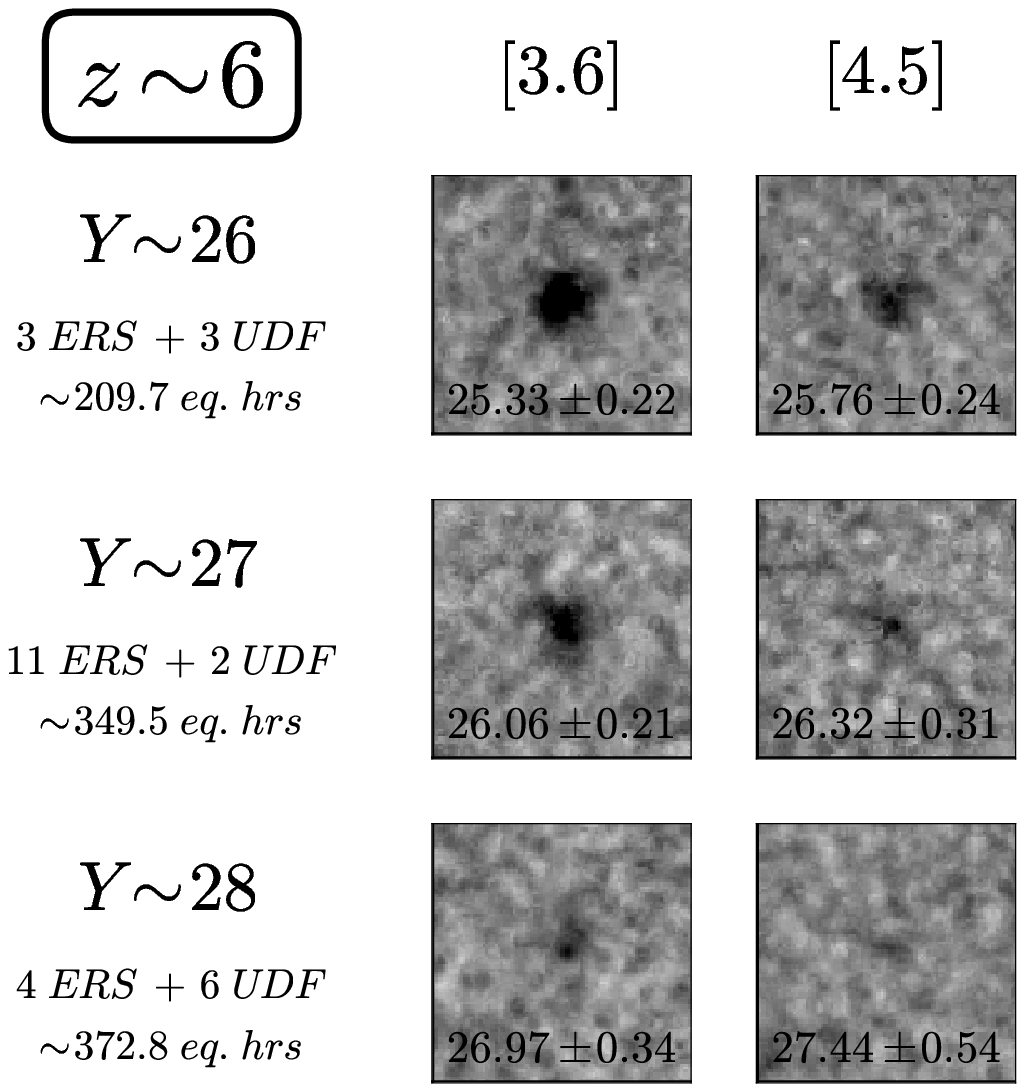}  
  \caption{As in Figure \ref{fig:BstackStamp} but for the $z\sim6$
    sample.  In this case the sample was split according to their
    $Y_{98}$ magnitude in the case of the ERS sources and $Y_{105}$
    for the HUDF sources (approximately rest-frame 1500\,\AA).}
  \label{fig:IstackStamp}
\end{#1}
}

\newcommand{\SEDsLee}[3]%
{
\begin{#1}[#2]
  \includegraphics[width=#3\textwidth]{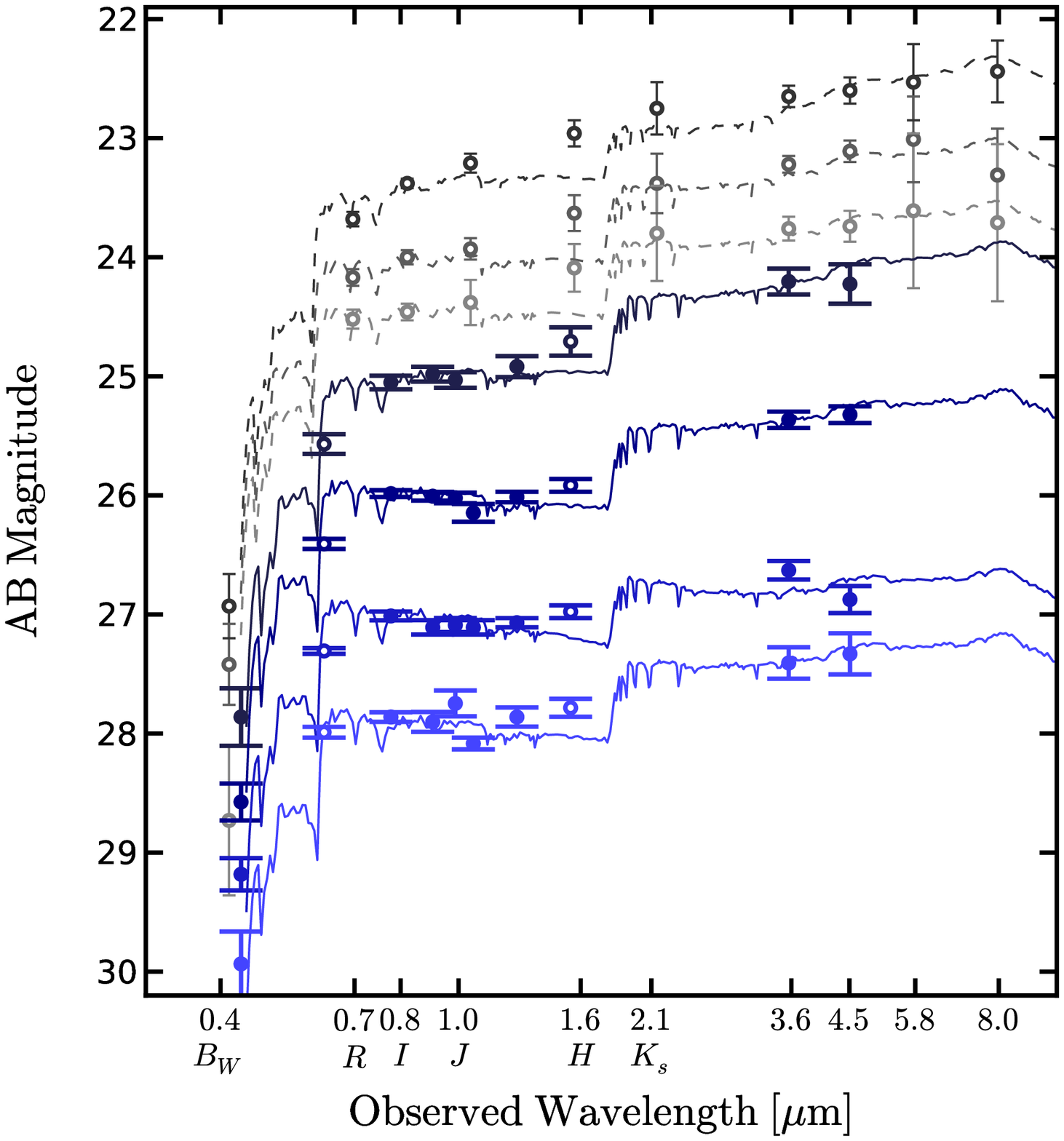}
  \caption{Stacked SEDs of $z\sim4$ galaxies in 7 bins of
  UV-luminosity. The faintest SEDs (blue solid symbols and lines)
  correspond to the SEDs derived from our sample.  This sample has a
  mean redshift of $\langle z\rangle=3.8\pm0.3$ and was split in bins
  of observed $\Delta i_{775} = 1\,$mag, approximately covering the
  UV-luminosity range $-21\lesssim M_{1500}\lesssim-18$.  The
  brightest stacks (gray open symbols and dashed lines) correspond to
  the median SEDs presented in \citet{lee11}, for a sample of galaxies
  with mean redshift $\langle z \rangle=3.7\pm0.4$. These SEDs allow
  us to extend the luminosity range to brighter luminosities up to
  $M_{1500}\sim-23$.  In \citet{lee11} the sources were split in six
  bins according to their observed $I$-band luminosity. For clarity,
  we only show every other bin but all are considered in Figure
  \ref{fig:LeebetaandUtoV}.  The x-axis shows wavelength and the
  approximate reference filter in their study.  Best fit BC03 models
  are included for references (the $H$ and $K_s-$band fluxes were
  excluded from the fits because they are likely biased -- see text,
  section 5.1). A trend of bluer UV-slopes and smaller Balmer breaks
  at fainter luminosities is observed.}
  \label{fig:LeeComparisonSEDs}
\end{#1}
}

\newcommand{\colorsLee}[3]%
{
\begin{#1}[#2]
  \includegraphics[width=#3\textwidth]{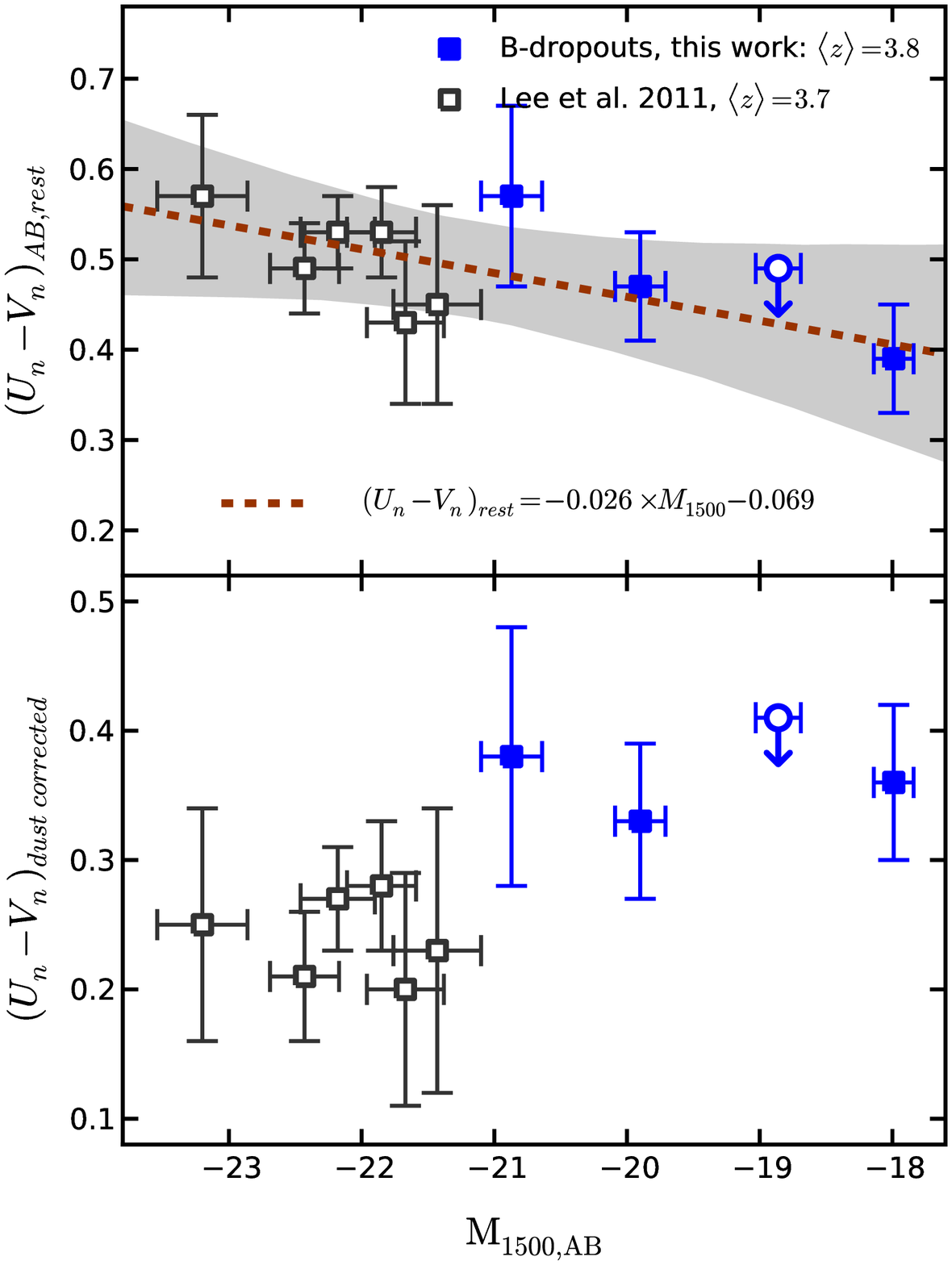}
  \caption{\emph{Top:} Interpolated rest frame $(U_n-V_n)$ color as a
  function of $M_{1500}$ for the $z\sim4$ sources.  Blue solid symbols
  and errorbars correspond to the colors derived for the new SEDs in
  this work.  Similarly, we have derived interpolated
  $(U_n-V_n)_{rest}$ colors for the SEDs presented in \cite{lee11},
  dark open symbols and errorbars.  Although not very steep, a
  systematic trend to bluer $(U_n-V_n)_{rest}$ colors with decreasing
  UV-luminosity is observed.  The best fit trend of $(U_n-V_n)_{rest}
  =-0.026 \times M_{1500} - 0.069$ is shown by the dashed line and the
  95\% confidence interval is shown by the gray area.  The rest-frame
  $(U_n-V_n)$ colors were determined from best fits to the observed
  data based on the default template set from the code EAzY
  \citep{bram08}.  We have used ideal narrow band $U_n$ and $V_n$
  filters that correspond to step functions of 100\,\AA\ width,
  centered at rest-frame 3500\,\AA\ and 5500\,\AA\ respectively. The
  upper limit ($2\,\sigma$), arises from the limitations of the
  template set which only includes $(U_n-V_n)>0.3$ mags.  The color of
  this SED is likely very close to this limit.  \emph{Bottom:} The
  same $(U_n-V_n)_{rest}$ colors after being corrected for dust
  extinction.  The corrections were derived from the median UV-slope
  $\beta$ trends from \citet{bouw09, bouw11a} and the local
  \citet{meur99} relation between $\beta$ and dust extinction. The
  corrected colors of both samples show flat trends with luminosity
  but seem to be disjoint. It is possible that the dust corrections
  are not equally adequate over the entire luminosity range. Another
  possible explanation is that the contamination rates are different
  between the two samples.  Finally, it should be noted that the
  UV-slope in the \citet{lee11} sample is only poorly sampled. Small
  offsets in the photometric calibration could bias the best fit
  models, from which the colors are estimated, to redder colors.}
  \label{fig:LeebetaandUtoV}
\end{#1}
}

\newcommand{\allSEDs}[3]%
{
\begin{#1}[#2]
  \includegraphics[width=#3\textwidth]{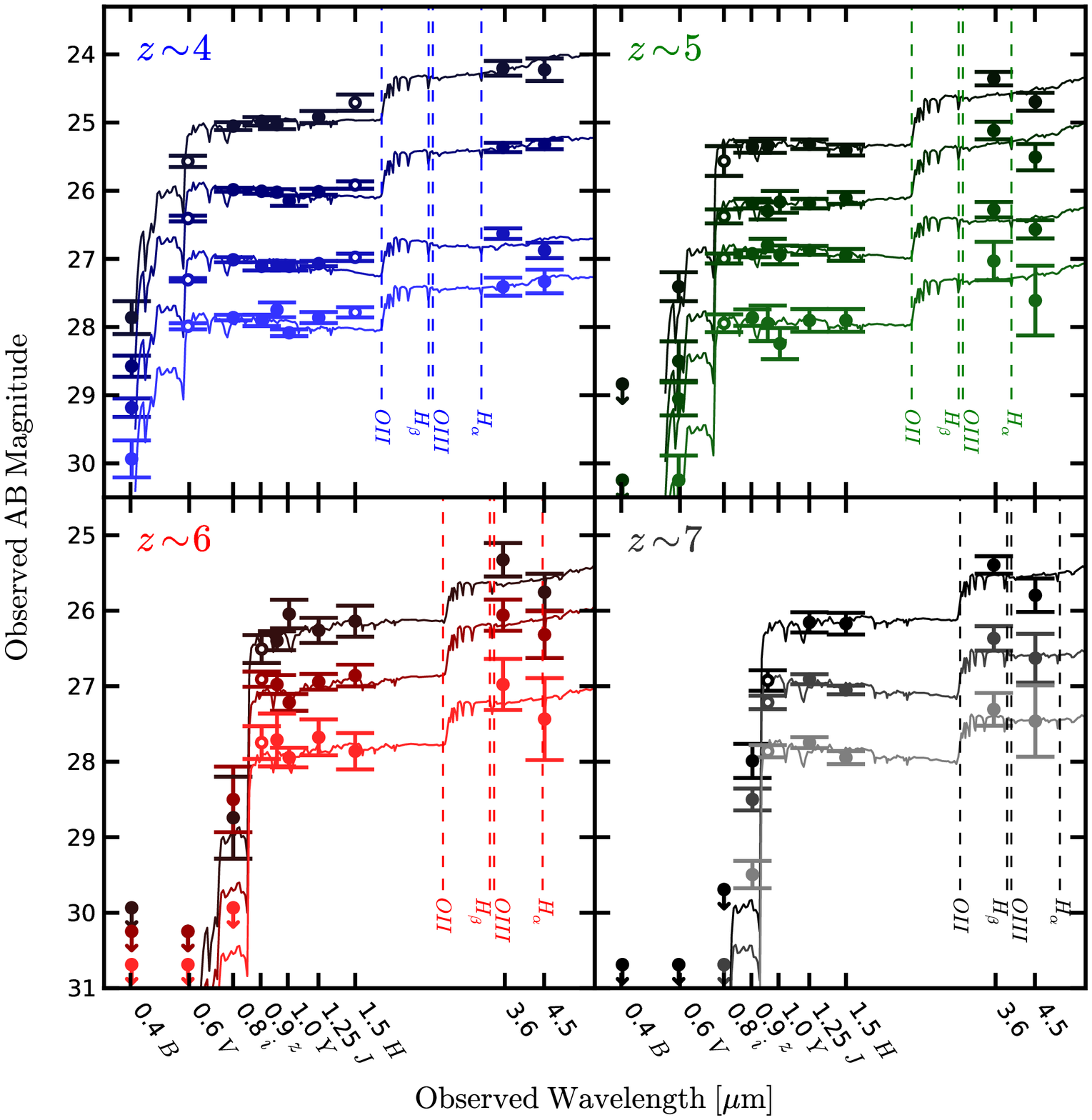}  
  \caption{The stacked SEDs of galaxies in units of observed
  magnitudes (see also Table \ref{tbl:SEDS}).  The x-axis shows
  wavelength and the filter that it corresponds to for reference. In
  the case of the optical bands, the errors were derived by bootstrap
  re-sampling the measured fluxes.  The individual uncertainties were
  set to a minimum of 5\% to account for systematic uncertainties in
  the filter to filter absolute calibrations. The errors in the IRAC
  bands were derived by bootstrap re-sampling the individual images
  and repeating the pixel by pixel median-stack process. In all cases,
  these errors on the median should include both the image noise and
  the uncertainty coming from variations within the population
  included in the stack.  Simple best fit BC03 models with CSF are
  also included for reference. In the fitting process the redshifts
  were fixed to the median redshift of each sample, i.e.: $\langle
  z\rangle=3.8$, $\langle z\rangle=5.0$, $\langle z\rangle=5.8$, and
  $\langle z\rangle=6.9$ respectively. Due to the intrinsic redshift
  distribution of each sample, the median fluxes measured for bands
  near Lyman alpha or the Balmer break are likely biased and in
  consequence they were excluded from the fitting process (excluded
  bands are marked by the open symbols). The position of the most
  prominent possible emission lines in the rest-frame optical region
  of the SED are marked by the vertical dashed lines (assuming the
  mean redshift of each sample).  At the different redshifts sampled
  here, different lines could be contaminating the continuum fluxes
  measured by the $Spitzer$/IRAC filters. The overall shape and
  UV-to-optical colors of the SEDs, however, remain remarkably
  constant with redshift. Some contribution of nebular emission lines
  is possibly hinted by the small relative excess of [3.6] over [4.5]
  fluxes observed at $z\gtrsim5$.}
  \label{fig:SEDs}
\end{#1}
}

\newcommand{\colorMag}[3]%
{
\begin{#1}[#2]
  \includegraphics[width=#3\textwidth]{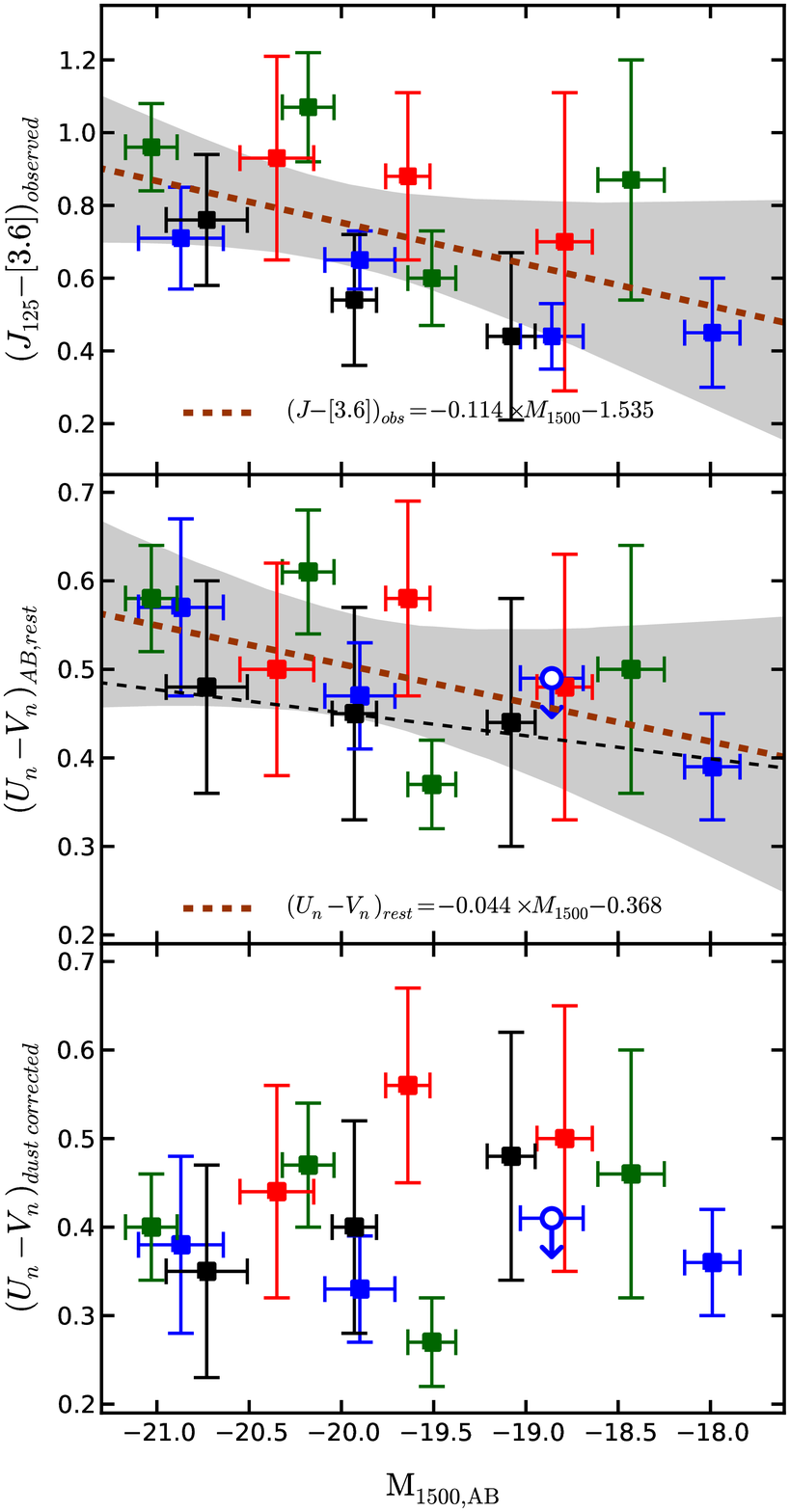}
  \caption{\emph{Top:} UV-to-optical color as a function of
  $M_{1500}$.  Blue symbols correspond to the $J-[3.6]$ colors of the
  $z\sim4$ SEDs; green: $z\sim5$; red: $z\sim6$; and black: $z\sim7$.
  Despite the sizable scatter, a systematic trend to redder
  UV-to-optical colors with increasing luminosity is observed at all
  redshifts.  At the same time, there is no clear indication of a
  systematic trend with redshift.  The dashed line is a best fit to
  all the points (all redshifts simultaneously): $J_{125} - [3.6] =
  -0.11(\pm0.07) \times M_{1500,AB}+const$, and the gray area
  corresponds to the 95\% confidence interval.  This is different to
  Figure \ref{fig:LeebetaandUtoV} in that this Figure shows the
  directly observed colors rather than interpolated rest-frame colors.
  In fact it is remarkable that this observed color, which samples
  different wavelength regions of the SED at the different redshift,
  remains fairly constant.  This is probably an indication that the
  rest-frame optical colors measured by IRAC are fairly flat.
  \emph{Center:} The interpolated rest frame $(U_n-V_n)$ colors
  (Section 5.1).  The best fit trend to all the data is shown by the
  brown dashed line.  For comparison, the black dashed line shows the
  best fit to the $z\sim4$ data only, as shown in Figure
  \ref{fig:LeebetaandUtoV}, this trend is slightly flatter.
  \emph{Bottom:} The $(U_n-V_n)_{rest}$ colors after being corrected
  by dust extinction following a simple prescription based on the
  UV-slopes and the local \citet{meur99} relation. Except for the
  $z\sim7$ SEDs (which have a much steeper $\beta$ vs. UV-luminosity
  relation), the dust corrected colors of all SEDs seem to show a flat
  trend (although the scatter is large).  This indicates that a change
  in dust is enough to explain the change in colors, both in the UV
  and the UV-to-optical.  In particular, no age dependence on
  luminosity seems to be required.} 
  \label{fig:colorMag}
\end{#1}
}

\newcommand{\combSEDs}[3]%
{
\begin{#1}[#2]
  \includegraphics[width=#3\textwidth]{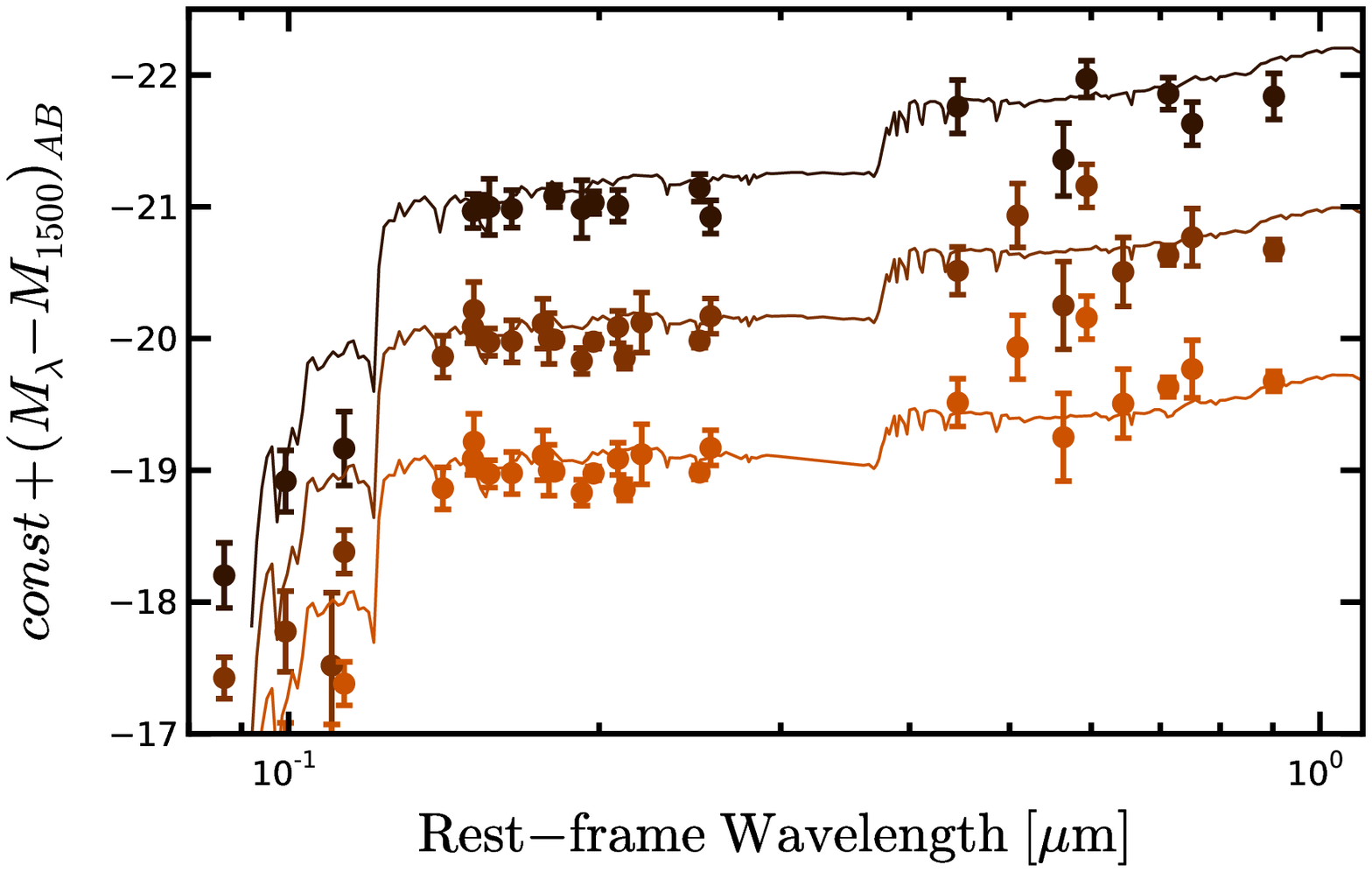}  
  \caption{The SEDs of the combined samples in three bins of intrinsic
  UV-luminosity: $M_{1500}\sim-21, -20,$ and -19. As seen in Figure
  \ref{fig:colorMag}, the UV-to-optical colors of the SEDs depend
  weakly on redshift.  Here we perform the exercise of combining all
  the photometry from SEDs at different redshifts into a single SED.
  At different redshifts, our set of HST/ACS+WFC3/IR+$Spitzer$/IRAC
  filters probe different rest-frame wavelengths.  The combined SEDs
  are plotted in terms of their magnitudes relative to $M_{1500}$,
  which is measured from the best power law fit to the UV continuum of
  each median SED.  An offset magnitude of $-21, -20,$ and $-19$ is
  added to reflect the approximate intrinsic magnitude of the SEDs
  that are being combined.}
  \label{fig:combSEDs}
\end{#1}
}

\newcommand{\figAgeDustModels}[3]%
{
\begin{#1}[#2]
  \begin{centering}
  \includegraphics[width=#3\textwidth]{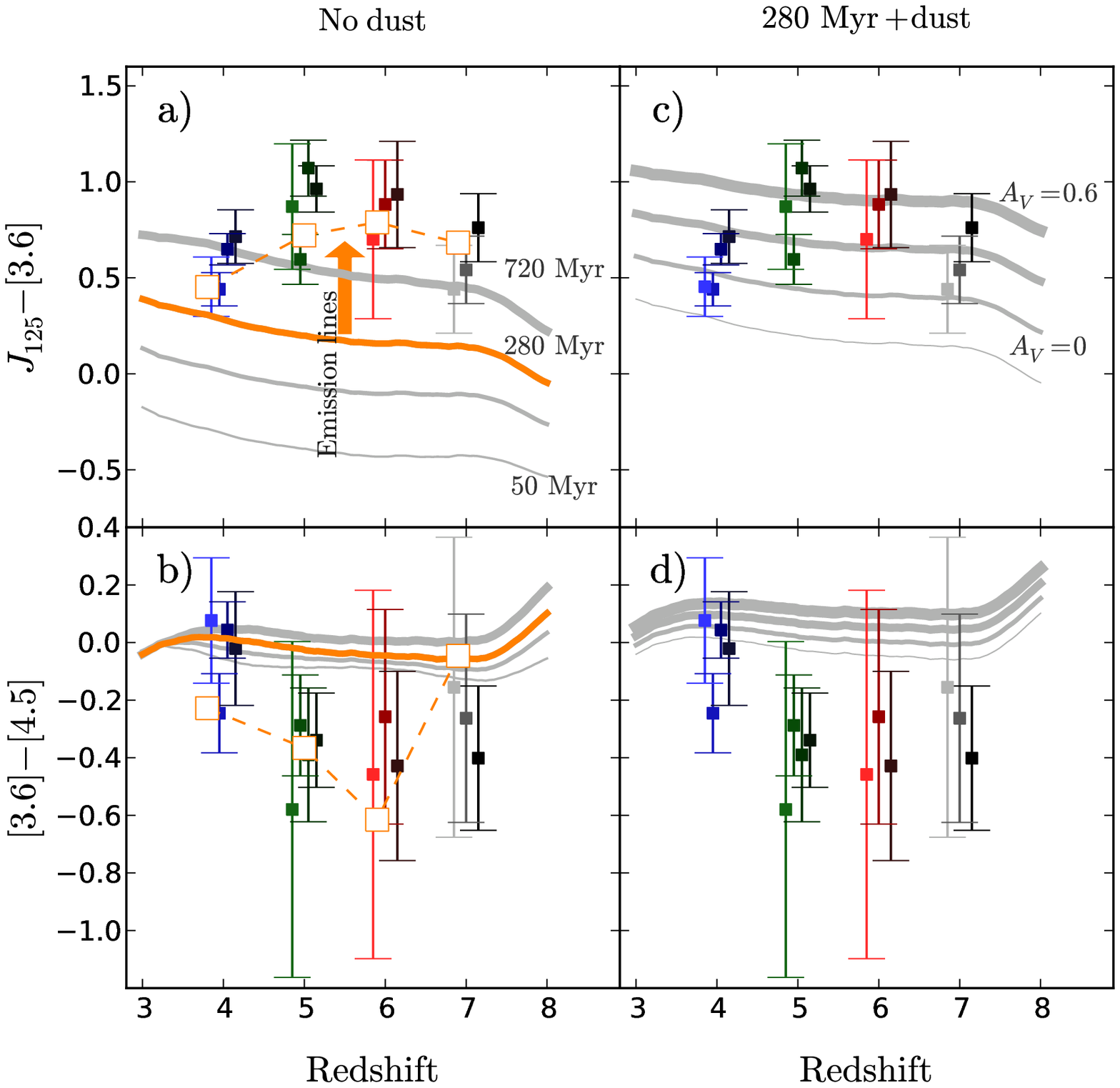}  
  \caption{The observed colors of the median SEDs.  The measurements
  are slightly offset in redshift for clarity. a) Theoretical tracks
  for a BC03 CSF model without dust are included.  The different lines
  correspond to ages: 50, 140, 280, and 720 Myr.  The 280 Myr old
  track is highlighted. For these dust-free models, the
  $J_{125}-[3.6]$ color is more consistent with older galaxy
  populations in which there is a significant contribution of
  relatively evolved stars to the rest-frame optical light. The most
  likely age is lowered if flux from optical nebular emission lines is
  included in the appropriate IRAC band and redshift. As an example of
  the changes that can result from the inclusion of emission lines,
  the open squares (connected by the dashed line) show the effects of
  adding emission lines with ${\rm EW_{rest}(OII, H\beta, OIII,
  H\alpha)=(189, 105, 670, 300)\,}$\AA, respectively to the 280\,Myr
  old track.  b) Dust free tracks (for the same ages as in 9a) cannot
  reproduce the $[3.6]-[4.5]\sim-0.3$ colors observed at $z\gtrsim5$.
  The inclusion of optical nebular emission lines, shown by the open
  squares connected by the dashed line can help reproduce such blue
  colors. c) As in a) but only the 280 Myr old track is shown with
  varying amounts of dust added. The tracks correspond, from bottom to
  top, to $A_V=0.0, 0.2, 0.4,$ and 0.6\,mag using a \citet{calz00}
  extinction law.  As expected, redder colors can be obtained for
  younger models if dust is allowed.  Our independent analysis of the
  UV continuum slope \citep{bouw11a}, constrains the amount of dust to
  $A_V<0.7\,$ mag for $z\gtrsim4$.  With this added constraint, the
  ages of high-$z$ galaxies are generally estimated to be $\gtrsim400$
  Myr when the emission line contribution is taken to be zero. d) As
  is clear here, the addition of dust makes it even harder for the
  stellar continuum only models to reproduce the blue $[3.6]-[4.5]$
  colors observed in the SEDs. This suggests that an optical nebular
  emission line component to the total flux may be needed.}
  \label{fig:ageModels}
  \end{centering}
\end{#1}
}
\newcommand{\tableSample}[1]%
{
\begin{deluxetable}{lcccc}[#1]
\tablecaption{\label{tbl:sampleSummary}Sample Summary.}
\tablehead{
  & \colhead{$z\sim4$} & \colhead{$z\sim5$} 
  & \colhead{$z\sim6$} & \colhead{$z\sim7\tablenotemark{a}$}
}
\startdata
ERS    &  270 / 524  &  77 / 123   &  20 / 32   &  15 / 18     \\
HUDF   &  137 / 205  &  41 / 55    &  38 / 51   &  21          \\
\tableline\\
TOTAL  &  407 / 729  &  118 / 178  &  58 / 83   &  36          \\
\enddata
\tablecomments{Summary of the number of sources in the sample.  IRAC
  photometry of these sources requires fitting and subtraction of the
  flux from surrounding foreground neighbors.  This is not possible in
  all the cases.  The table shows the number of sources with clean
  IRAC photometry / total number of sources found in the field.}
\tablenotetext{a}{We adopt the sample presented in
  \citealt{labb10}.}
\end{deluxetable}
}

\newcommand{\tablePhotometry}%
{
\begin{turnpage}
\begin{deluxetable*}{cccrrrrrrrrrr}[h!]
\tablecaption{\label{tbl:SEDS}Photometry Summary of Stacked SED.}
\tablehead{
  \colhead{Ref. mag} & \colhead{$N_{ERS}$} & \colhead{$N_{HUDF}$} &
  \colhead{$B_{435}$} & \colhead{$V_{606}$} & \colhead{$i_{775}$} &
  \colhead{$z_{850}$} & \colhead{$Y_{98}$}  & \colhead{$Y_{105}$} &
  \colhead{$J_{125}$} & \colhead{$H_{160}$} &
  \colhead{$[3.6]$}  & \colhead{$[4.5]$}
}
\startdata
$i_{775}$&\multicolumn{12}{c}{$z\sim4 ~ B-$dropouts}\\
25 &  27 / 55 &  0 / 4  & $27.5\pm7.2$ & $224.0\pm14.9$ & $363.2\pm18.8$ & $384.8\pm24.7$ & $366.5\pm22.4$ & \nodata       & $405.5\pm33.0$ & $480.7\pm53.4$ & $784.5\pm78.3$ & $768.6\pm117.2$\\ 	
26 &  87 / 16 & 12 / 17 & $13.5\pm2.4$ & $103.1\pm 3.8$ & $150.1\pm 4.1$ & $149.1\pm 5.8$ & $146.8\pm 5.9$ & $130.6\pm8.9$ & $147.9\pm 5.3$ & $161.8\pm 7.1$ & $268.8\pm17.1$ & $280.0\pm 17.6$\\ 	
27 & 107 / 2  & 25 / 34 & $ 8.4\pm1.2$ & $ 45.5\pm 1.7$ & $ 58.9\pm 1.8$ & $ 54.4\pm 2.4$ & $ 55.0\pm 2.9$ & $ 53.1\pm3.2$ & $ 56.1\pm 2.3$ & $ 60.9\pm 3.4$ & $ 83.8\pm 6.3$ & $ 67.0\pm  7.5$\\ 	
28 &  45 / 78 & 43 / 74 & $ 3.6\pm1.0$ & $ 24.4\pm 1.0$ & $ 27.2\pm 1.0$ & $ 26.0\pm 1.6$ & $ 29.4\pm 3.3$ & $ 22.4\pm1.5$ & $ 26.8\pm 1.5$ & $ 29.0\pm 2.1$ & $ 41.4\pm 4.9$ & $ 43.7\pm  7.2$\\ 	
\tableline\\
$z_{850}$&\multicolumn{12}{c}{$z\sim5 ~ V-$dropouts}\\
25 & 4 / 9    & 0 / 0  & $-6.2\pm11.7$ & $42.1\pm8.1$ & $227.4\pm40.8$ & $267.8\pm21.8$ & $273.0\pm22.6$ & \nodata        & $281.9\pm18.4$ & $258.3\pm19.3$ & $681.9\pm62.2$ & $498.8\pm59.9$\\
26 & 18 / 28  & 2 / 3  & $ 0.9\pm 3.0$ & $15.9\pm3.7$ & $104.7\pm11.9$ & $125.6\pm 8.1$ & $114.5\pm13.2$ & $128.4\pm17.4$ & $126.3\pm 8.7$ & $137.3\pm11.1$ & $337.9\pm40.3$ & $236.5\pm41.5$\\
27 & 43 / 61  & 8 / 9  & $-2.9\pm 2.1$ & $ 8.3\pm1.7$ & $ 58.9\pm 4.2$ & $ 63.9\pm 3.0$ & $ 70.0\pm 5.6$ & $ 62.1\pm 8.0$ & $ 67.7\pm 3.9$ & $ 63.0\pm 4.6$ & $116.3\pm11.5$ & $ 89.4\pm10.6$\\
28 & 11 / 24  & 9 / 13 & $-0.8\pm 1.9$ & $ 3.5\pm1.4$ & $ 24.5\pm 3.0$ & $ 26.8\pm 2.6$ & $ 25.5\pm 6.2$ & $ 19.0\pm 3.7$ & $ 25.9\pm 3.9$ & $ 26.5\pm 3.9$ & $ 58.0\pm15.4$ & $ 33.8\pm15.7$\\
\tableline\\
$Y$     &\multicolumn{12}{c}{$z\sim6 ~ i-$dropouts}\\
26 & 3 / 6   & 3 / 3   & $ 1.2\pm3.8$ & $-1.5\pm3.6$ & $12.1\pm6.6$ & $93.0\pm14.7$ & $102.2\pm12.1$ & $143.2\pm23.8$ & $116.4\pm14.8$ & $133.6\pm26.3$ & $278.7\pm56.8$ & $187.7\pm42.1$\\
27 & 11 / 17 & 2 / 3   & $-0.5\pm3.4$ & $ 0.1\pm2.4$ & $14.2\pm5.0$ & $64.5\pm 6.5$ & $ 60.2\pm 7.4$ & $ 48.7\pm 5.1$ & $ 64.2\pm 5.9$ & $ 68.4\pm 7.8$ & $142.2\pm27.0$ & $111.9\pm31.8$\\
28 & 4 / 4	 & 6 / 13  & $-2.5\pm2.5$ & $ 0.5\pm2.1$ & $-0.7\pm3.6$ & $30.3\pm 5.4$ & $ 30.7\pm 8.9$ & $ 24.6\pm 3.4$ & $ 33.0\pm 5.8$ & $ 26.2\pm 6.3$ & $ 61.3\pm19.2$ & $ 39.9\pm20.0$\\
\tableline\\
$H_{160}$&\multicolumn{12}{c}{$z\sim7 ~ z-$dropouts}\\
26 & \nodata & \nodata & $-3.2\pm2.3$ & $ 2.8\pm2.3$ & $-7.4\pm4.7$ & $23.5\pm5.4$ & $63.7\pm7.7$ & \nodata & $129.7\pm16.0$ & $128.2\pm17.0$ & $262.0\pm28.0$ & $181.0\pm37.0$\\
27 & \nodata & \nodata & $-2.4\pm0.8$ & $-0.8\pm0.6$ & $ 2.2\pm1.8$ & $14.6\pm2.3$ & $49.1\pm4.3$ & \nodata & $ 64.8\pm 4.1$ & $ 57.0\pm 3.4$ & $107.0\pm16.0$ & $ 83.8\pm25.0$\\
28 & \nodata & \nodata & $ 1.2\pm0.8$ & $ 0.4\pm0.5$ & $-0.1\pm0.7$ & $ 5.6\pm1.1$ & $27.2\pm1.8$ & \nodata & $ 29.6\pm 2.0$ & $ 24.5\pm 1.6$ & $ 45.1\pm 9.5$ & $ 39.3\pm17.0$\\
\enddata

\tablecomments{The median SEDs of high$-z$ galaxies in bins of
  observed UV-luminosity.  The fluxes are in units of
  nJy.  $N_{HUDF}$ and $N_{ERS}$ are the numbers of sources with
  reliable IRAC photometry over the total available in the given
  bin. Only the ones with reliable photometry were considered in the
  stack. The $z\sim7$ stacks were taken directly from \citet{labb10}.}
\end{deluxetable*}
\end{turnpage}
}

\newcommand{\tableColors}%
{
\begin{deluxetable}{cccc}[h!]
\tablecaption{\label{tbl:colors}Rest Frame Magnitude and Colors.}
\tablehead{
  \colhead{$M_{1500}$} & \colhead{$J_{125}-[3.6]$} &
  \colhead{$(U-V)_{rest}$} & \colhead{$(U-V)_{dust-corr}$}
}
\startdata
\multicolumn{4}{c}{$z\sim4 ~ B-$dropouts}\\
$-20.87\pm0.23$ & $0.71\pm0.14$ & $0.57\pm0.10$ & $0.38$\\ 
$-19.90\pm0.19$ & $0.65\pm0.08$ & $0.47\pm0.06$ & $0.33$\\ 
$-18.86\pm0.17$ & $0.44\pm0.09$ & $ > 0.49$    & $>0.41$\\ 
$-17.99\pm0.15$ & $0.45\pm0.15$ & $0.39\pm0.06$ & $0.36$\\ 
\tableline\\
\multicolumn{4}{c}{$z\sim5 ~ V-$dropouts}\\
$-21.03\pm0.14$ & $0.96\pm0.12$ & $0.58\pm0.06$ & $0.40$\\ 
$-20.18\pm0.14$ & $1.07\pm0.15$ & $0.61\pm0.07$ & $0.47$\\ 
$-19.51\pm0.13$ & $0.60\pm0.13$ & $0.37\pm0.05$ & $0.27$\\ 
$-18.43\pm0.18$ & $0.87\pm0.33$ & $0.50\pm0.14$ & $0.46$\\ 
\tableline\\
\multicolumn{4}{c}{$z\sim6 ~ i-$dropouts}\\
$-20.35\pm0.20$ & $0.93\pm0.28$ & $0.50\pm0.12$ & $0.44$\\ 
$-19.64\pm0.12$ & $0.88\pm0.23$ & $0.58\pm0.11$ & $0.56$\\ 
$-18.79\pm0.15$ & $0.70\pm0.41$ & $0.48\pm0.15$ & $0.50$\\ 
\tableline\\
\multicolumn{4}{c}{$z\sim7 ~ z-$dropouts}\\
$-20.73\pm0.22$ & $0.76\pm0.18$ & $0.48\pm0.12$ & $0.35$\\ 
$-19.93\pm0.12$ & $0.54\pm0.18$ & $0.45\pm0.12$ & $0.40$\\ 
$-19.08\pm0.13$ & $0.44\pm0.23$ & $0.44\pm0.14$ & $0.48$\\ 
\enddata
\tablecomments{Rest frame $M_{1500}$ estimated from the best fit
power-law to the rest-frame UV photometry. Possibly contaminated bands
were ignored in the fit (see open points in Figure \ref{fig:SEDs}).}
\end{deluxetable}
}

\begin{document}
\title{The Rest Frame UV to Optical Colors and SEDs of $z\sim4-7$ galaxies.}

\author{
Valentino Gonz\'alez\altaffilmark{1},
Rychard J. Bouwens\altaffilmark{2},
Ivo Labb\'e\altaffilmark{2},
Garth Illingworth\altaffilmark{1},
Pascal Oesch\altaffilmark{1,3},
Marijn Franx\altaffilmark{2},
Dan Magee\altaffilmark{1}
}

\altaffiltext{1}{Astronomy Department, University of California,
Santa Cruz, CA 95064}    
\altaffiltext{2}{Leiden Observatory, Leiden University , NL-2300 RA
Leiden, Netherlands} 
\altaffiltext{3}{Hubble Fellow}

\begin{abstract}

  We use the ultra-deep HUDF09 and the deep ERS data from the HST
  WFC3/IR camera, along with the wide area $Spitzer$/IRAC data from
  GOODS-S to derive SEDs of star-forming galaxies from the rest-frame
  UV to the optical over a wide luminosity range  ($M_{1500}\sim -21$
  to $M_{1500}\sim-18$) from $z\sim7$ to $z\sim4$. The sample contains
  $\sim 400\ z\sim4$, $\sim 120\ z\sim 5$, $\sim 60\ z\sim 6$, and 36
  prior $z\sim7$ galaxies. Median stacking enables the first
  comprehensive study  of very faint high-z galaxies at multiple
  redshifts (e.g., $[3.6] = 27.4 \pm 0.1$ AB mag for the
  $M_{1500}\sim-18$ sources at $z\sim4$).  At $z\sim4$ our faint
  median-stacked SEDs reach to $\sim0.06L^*_{z=4}$ and are combined
  with recently published results at high luminosity $L>L^*$ that
  extend to $M_{1500}\sim-23$. We use the observed SEDs and template
  fits to derive rest frame UV-to-optical colors $(U - V)$ at all
  redshifts and luminosities. We find that this color does not vary
  significantly with redshift at a fixed luminosity.  The
  UV-to-optical color does show a weak trend with luminosity, becoming
  redder at higher luminosities.  This is most likely due to dust.  At
  $z\gtrsim5$ we find blue colors $[3.6]-[4.5]\sim-0.3\,$mag that are
  most likely due to rest-frame optical emission lines contributing to
  the flux in the IRAC filter bandpasses. The scatter across our
  derived SEDs remains substantial, but the results are most
  consistent with a lack of any evolution in the SEDs with redshift at
  a given luminosity. The similarity of the SEDs suggests a
  self-similar mode of evolution over a timespan from 0.7 Gyr to 1.5
  Gyr that encompasses very substantial growth in the stellar mass
  density in the universe (from $\sim4\times10^6$ to $\sim2\times10^7$
  ${\rm M_\odot Mpc^{-3}}$)

\end{abstract}

\keywords{galaxies: evolution --- galaxies: high-redshift}

\section{Introduction}

As a result of the installation of the WFC3/IR camera on the Hubble
Space Telescope, it is now possible to obtain extremely deep, high
resolution observations in the near infrared. These observations match
the exquisite optical data provided by the ACS camera over deep and
ultra-deep fields like the GOODS and the HUDF \citep{giav04, bouw07}.
The upgrade has enabled several teams to extend their searches for
Lyman Break Galaxies (LBGs) to very high redshifts
\citep[e.g.,][]{bouw10b, oesc10, oesc11, mclu10, yan10, bunk10,
fink10}. In combination, these two cameras have allowed for high S/N
measurements of the rest-frame UV light of high redshift galaxies over
a wide wavelength baseline. 

As a result of these new observations, the UV LFs at $z\gtrsim4$ are
known to high accuracy. Another direct result from these observations
is the determination of the UV slope of the SEDs, generally
characterized by the parameter $\beta$
($f_\lambda\propto\lambda^\beta$, e.g., \citealt{meur99, bouw09}).
This slope has been shown to become steeper (bluer colors) both with
decreasing luminosity and with increasing redshift (\citealt{bouw09,
bouw11a, wilk11}, but see \citealt{dunl11, fink11}). For a mid-aged
($100\,$Myr) stellar population, this slope generally depends most
strongly on the dust reddening so this has been interpreted as an
increase in the dust content of galaxies with cosmic time and as
galaxies increase in luminosity \citep{bouw11a, fink11}.

Since the rest-frame UV light primarily probes young short-lived
stellar populations (mostly O and B stars), it is interesting to
extend these observations to longer wavelengths and study the relative
importance of more evolved stellar populations.  This is very
challenging though and is now only possible thanks to the incredible
sensitivity of the IRAC camera on the $Spitzer$ Space Telescope. Some
of the first observations of rest-frame optical light from $z\gtrsim4$
galaxies showed quite substantial UV-to-optical colors \citep{eyle05,
yan05}.  Under the usual assumption of smooth, exponentially declining
star formation histories (SFHs), these colors were interpreted as
indicative of an evolved stellar population (ages $\gtrsim100\,$ Myr)
in combination with the younger populations probed by the UV-light.
Follow-up work on $z\sim4-7$ galaxies continued to reveal systems with
moderate size UV-to-optical breaks \citep[e.g.][]{eyle07, yan06,
labb06, star09, gonz10}, even possibly to $z\sim8$ \citep{labb10}.
However interesting, these studies are only possible for the brightest
$z>4$ galaxies, due to the extreme depths required.

Some alternative explanations have been proposed for the red
UV-to-optical colors observed in $z>6$ galaxies.  For example, for
young star-froming galaxies, one might expect strong line emission.
At the right redshift, these lines could contaminate the 3.6 and 4.5
$\mu m$ IRAC fluxes, resulting in red UV-to-optical colors similar to
those produced by the Balmer break \citep{scha09, scha10}.  For
emission lines to be strong enough to dominate the rest-frame optical
fluxes, extremely young ages ($\sim6\,$Myr) are required. Such models
have ages shorter than the dynamical times of these high$-z$ sources
and are also somewhat uncertain. It would be surprising if the
majority of these galaxies, which are selected over a redshift range
that spans $\gtrsim200\,$Myr, had such young ages. Even if rest-frame
optical emission nebular emission lines do not dominate the measured
IRAC fluxes, it is expected that they will contribute some of the flux
measured in the mid-IR, but the exact contribution currently remains
hard to establish at these high redshifts.

It has been routine, then, to compare the colors of these high
redshift galaxies with synthetic stellar population (SSP) models of
relatively evolved galaxies with smooth SFHs to derive physical
properties from the observations.  By studying LBG samples at
different redshifts in the aforementioned way, one intriguing result
has emerged.  At a given intrinsic luminosity, the specific Star
Formation Rate (SFR) of galaxies seems to remain remarkably constant
with redshift (\citealt{star09,gonz10, gonz11, mclu11}; slightly
modified to account for the latest dust corrections by
\citealt{bouw11a}).  This result has called the attention of theorists
and has proved very hard to reproduce in simulations
\citep[e.g.,][]{wein11, khoc11}.  The assumptions used to derive these
result vary somewhat from group to group.  In this work, we present
the basic observations that lead to such a result, namely the observed
SEDs of galaxies at $z\sim4-7$.

In the present work we study the observational properties of the
typical high redshift galaxy.  We do this by splitting our large
sample in both redshift and UV-luminosity bins and constructing mean
SEDs.  The large number of galaxies allows us to study the typical
colors at high S/N but also extend our measurements to faint limits
that are currently inaccessible on an individual basis with the
current data.  Although the scatter about the typical properties is
also highly interesting, we defer their study to a future work which
relies on even deeper $Spitzer$ observations.  In section 2 we present
the Data and in section 3 the sample selection and photometry. Section 4
describes the stacking procedure and we discuss the stacked SEDs and
colors in section 5.  A summary is presented in section 6.

Throughout, we use a ($H_0,\,\Omega_M,\,\Omega_\Lambda$) = ($70\,\rm{km
  ~s^{-1}}$, 0.3, 0.7) cosmology when necessary and we quote all
magnitudes in the AB system \citep{oke83}.

\tableSample{h!}
\figMagHist{figure*}{ht!}{0.75}

\section{Data} 

In this paper, we make use of the large number of $z\gtrsim4$ galaxies
found in the ultra-deep HUDF data and wide-area ERS data to construct
median SEDs for star-forming galaxies at $z\sim4-7$.   The deep
optical, near-IR, and deep IRAC coverage over these fields allow us to
create SEDs that cover both the rest-frame UV and the rest-frame
optical down to very faint fluxes.

\subsection{HST ACS and WFC3/IR}

In constructing our median SEDs, we make use of the deep optical and
near-IR HST data from ACS and WFC3/IR.  Over the ERS field, both the
ACS optical ($B_{435}$$V_{606}$$i_{775}$$z_{850}$) and the WFC3/IR
($Y_{98}$$J_{125}$$H_{160}$) images from HST reach depths of $\sim28$
mag ($5\,\sigma$ measured on 0.35\arcsec-diameter apertures).  In the
HUDF field, the ACS optical data ($BViz$) are $1.5-2$ mags deeper, and
the WFC/IR data ($Y_{105}J_{125}H_{160}$) are 1.5 times deeper than
the ERS.  The HUDF was observed with the $Y_{105}$ filter instead of
the $Y_{98}$ filter used in the ERS.

\subsection{Spitzer/IRAC}

Because these fields are located inside the GOODS-S field, they both
have deep $Spitzer$/IRAC coverage that amounts to $\sim23.3\,$hr over
the ERS and twice that for the HUDF which was observed to the same
depths but in two different epochs with the IRAC camera rotated by
$180^{\circ}$. For IRAC channel 1 images ($\lambda_\circ=3.6\mu m$),
each single epoch reaches depths of 27.8 mags, measured over
2.5\arcsec-diameter apertures ($1\,\sigma$).  For channel 2
($\lambda_\circ=4.5\mu m$) this limit corresponds to 27.2 mags.  

\section{Sample Selection and Basic Photometry}

The criteria used to identify high redshift star forming galaxies has
already been presented in numerous previous works. For details on the
selection procedure and discussion of sample contamination please see
\citet{bouw07, bouw10d}.  In short, the selection criteria used to find
the sources presented here is as follows:\\

$z\sim4~B$-dropouts:
$$(B_{435}-V_{606}>1.1)~\wedge~[B_{435}-V_{606}>(V_{606}-z_{850})+1.1]$$
$$\wedge~(V_{606}-z_{850}<1.6)$$

$z\sim5~V$-dropouts:
$$\{[V_{606}-i_{775}>0.9(i_{775}-z_{850})]~\vee~(V_{606}-i_{775}>2)\}$$
$$\wedge~(V_{606}-i_{775}>1.2)~\wedge~(i_{775}-z_{850}<1.3)$$

$z\sim6~i$-dropouts:
$$(i_{775}-z_{850}>1.3)~\wedge~(z_{850}-J_{125}<0.8)$$

The combined samples contain a total of 729 sources at $z\sim4$, 178
at $z\sim5$, and 83 at $z\sim6$ (see table \ref{tbl:sampleSummary}).

We complement the $z\sim4,~5$, and 6 samples with the sample of
36 $z\sim7$ sources presented by \citet{labb10}. We have adopted the
$z\sim7$ stacks presented in that work.

At these depths, the $Spitzer$/IRAC images are fairly crowded due to
the size and shape of its PSF.  As a consequence, standard aperture
photometry of the sources generally results in fluxes that are
contaminated by flux coming from unrelated sources that are nearby in
the sky. We are able to de-blend the fluxes from the different sources
by making use of the higher resolution information available through
the HST images.  Our method has been explained in previous works by
our group \citep[e.g][]{labb06, labb10a, gonz10, gonz11} but we give a
short description here.  We start by registering the IRAC images to
the higher resolution HST image that will be used as a light profile
template. In this case we use the WFC3/IR $H_{160}$-band image as our
template because it is the closest in wavelength and has high signal
to noise.  We use the SExtractor software \citep{bert96} to create
segmentation maps. We isolate all the sources in a small area around
each galaxy in our sample using these segmentation maps. We then
convolve them with a kernel derived from the PSFs of the HST and the
IRAC images. We fit for the normalization fluxes of all sources
(including the source we are trying to clean) simultaneously.  Finally
we remove the flux from the unrelated sources to obtain a clean image
of the galaxies in our sample.  Later we use these cleaned images to
produce median stacks.

The procedure described above does not always produce acceptable
results. The reasons why it can fail vary but the most common is the
presence of a source that is too bright, extended, and close to our
region of interest.  In these cases the amount of flux subtracted off
our source of interest is large and more uncertain and in most cases
results in large residuals.  We have carefully inspected the residual
images (all sources subtracted, including the one in our sample) and
the cleaned images (only unrelated neighbors subtracted) of each of
the sources in the original sample and come up with a sub-sample of
sources with clean IRAC photometry.  This ``by eye'' inspection
criterion generally results in good agreement with a criterion that
selects only the images with the lowest $\chi^2$ residuals but allows
us to also remove sources that, even though have moderate $\chi^2$
values, present mild to strong residuals located very close to the
source. Overall, this reduces the sample to 60\% of the original (see
Table \ref{tbl:sampleSummary}).  Since the criterion adopted here
depends on the complexity of the unrelated sources, it is not expected
that it should introduce any biases that would be relevant in the
determination of the mean rest-frame optical colors (see Figure
\ref{fig:magDist}).

\section{Stacked Photometry and Median SEDs}

It has been shown that in star forming galaxies the total SFR appears
to correlate with stellar mass, the so called main sequence of star
forming galaxies \citep{noes07}. This sequence has been observed at
high-redshifts ($z\gtrsim4$) but contrary to what has been observed at
$z<2$, the normalization of the relation at fixed stellar mass does
not seem to evolve over $2 < z < 7$ \citep{star09, gonz11}. The lack
of deeper rest-frame optical from $Spitzer$/IRAC for these sources is
one of the big limitations to study this sequence to fainter
magnitudes.  In terms of the bulk of the population, however, progress
can be made by stacking large numbers of sources to study their mean
properties.  In this section we seek to construct the median SEDs of
galaxies at $z\gtrsim4$.

Due to the nature of the dropout search and the filters available, our
full sample is naturally divided in redshift bins centered at $\langle
z\rangle=3.8, 5.0, 5.9$ plus the $\langle z\rangle=6.9$ sample from
\citet{labb10}.  In this section, we further split each redshift
sub-sample based on their observed rest-frame UV-luminosity.  The
filters used as reference UV-luminosity are the $i_{775}$ filter for
the $z\sim4$ sources; the $z_{850}$ for the $z\sim5$ sources; the
$Y_{98}$ for the $z\sim6$ sources in the ERS and the $Y_{105}$ for the
$z\sim6$ sources in the HUDF; and finally, the $J_{125}$ filter the
$z\sim7$ sources.  These filters were chosen to be close to rest-frame
1500\,\AA\  at the different redshifts and make the splitting criterion
fairly uniform across redshifts.  These bins are presented in Figure
\ref{fig:magDist}.

In the case of the HST ACS and WFC3 imaging, the fluxes utilized
correspond to the MAG\_AUTO values as determined using the SExtractor
code.  For each UV-luminosity bin, the median flux in each band was
determined from the individual flux measurements of the sources in the
bin that also have clean IRAC determinations.  The uncertainty in the
median was estimated through a bootstrap process. In each realization,
$N_{bin}$ fluxes are drawn with replacement from the $N_{bin}$ values
available and perturbed by their individual uncertainties.  Generally
the individual HST flux determinations have very high signal to noise.
There are however, systematic uncertainties in the absolute
calibration of the different bands and instruments across the SED.
These uncertainties amount to at least 5\% which we adopt as our
minimum allowed uncertainty in any given band.

In the case of the IRAC imaging, most sources are too faint to be
individually detected even in these deep $\sim\,23.3\,$hr integration
images. Instead of using individual fluxes, we first median combine
(pixel by pixel) the cleaned stamps of the sources within a bin that
have clean residuals from our IRAC cleaning procedure. We perform
standard aperture photometry on the stacked images using a circular
aperture of 2.5\arcsec-diameter and correct the fluxes to total
assuming PSF profiles. This amounts to a factor 1.8 in both the [3.6]
and the [4.5] IRAC filters. The local background is estimated from the
stacks, in a ring between 6\arcsec and 10\arcsec diameters around the
stamp center, where the source is expected to be, and then subtracted
off the image.

To estimate the uncertainty in the IRAC stacks, we also create
bootstrap realizations. We treat the two epochs of IRAC data over the
HUDF as independent stamps in the stack, thereby effectively weighting
with exposure time, i.e., $N_{bin}=N_{ERS}+2N_{HUDF}$. In each
bootstrap draw, then, we draw with replacement $N_{ERS}+2N_{HUDF}$
such stamps, and estimate fluxes as described above.  The image noise
in each stacks is expected to decrease with the total equivalent
exposure time approximately as $1/\sqrt{t_{exp,eq}}$
($t_{exp,eq}=(N_{ERS}+2N_{HUDF})\times23.3\,$hr).  The actual
photometric uncertainties that we derive are always greater than this
because they also include the intrinsic scatter in the fluxes of the
sources been stacked together (see Appendix).  Stamps for all the IRAC
median stacks along with the derived magnitudes and uncertainties are
presented in Figures \ref{fig:BstackStamp}, \ref{fig:VstackStamp}, and
\ref{fig:IstackStamp}. Table \ref{tbl:SEDS} is a compilation of all
the stacked photometry including the \citet{labb10} stacks and
information on the number of sources that go in each stack. It can be
seen in the Figures that even for sources as faint as 27\,mag, the
stacks show good detections in both IRAC bands.

\Bstamps{figure}{h}{0.4}
\Vstamps{figure}{h}{0.4}
\istamps{figure}{h}{0.4}

\section{Rest-Frame UV-to-optical Colors of high-$z$ Star-Forming Galaxies.}

The observed colors of galaxies can give us important information
about the stellar populations that compose them.  Usually, this is
accomplished by comparing their colors with those of Synthetic Stellar
Population (SSP) models.  In this work, we chiefly explore the
observed UV-to-optical colors and what can be learned from them and
from their dependence on luminosity and redshift. When models are
invoked, we use the \citet[BC03]{bruz03} models and we assume a
\citet{salp55} IMF, constant SFH, and $0.2\,Z_\odot$ metallicity, as
these quantities cannot be constrained from our data. To derive best
fits we use the SED fitting code FAST \citep{krie09}.

The rest-frame UV-to-optical color is of particular interest at high
redshifts.  Several studies have detected significant rest-frame
UV-to-optical colors ($\gtrsim0.5\,$mag) in $z\gtrsim4$ galaxies
\citep{eyle05, eyle07, yan05, star09, gonz10, gonz11, labb06, labb10,
mclu11}. The usual interpretation is that these colors derive from
large Balmer breaks which are associated with evolved stellar
populations with ages $\gtrsim200\,$Myr. This is surprising given how
young the universe is at the time these galaxies are being observed.
The main limitation for field studies is the depth of the
$Spitzer$/IRAC imaging available, which constrains these studies to
the brightest LBGs.  Searches behind clusters provide an alternative
to probe the fainter populations (see, e.g., \citealt{elli01, brad08,
rich08, lapo11}, and the recent tentative determination of a 1.5 mag
Balmer break in a $z\sim6$ galaxy by \citealt{rich11}) but the number
of sources amenable to such studies, remain low.  We work around this
limitation by stacking large numbers of sources which allows us to
estimate the typical rest-frame UV-to-optical colors of faint
$z\gtrsim4$ sources.  

We start by presenting the results found at $z\sim4$ where the similar
and complementary work of \citet{lee11} allows us to study this colors
over an unprecedentedly large range of UV-luminosities.

\SEDsLee{figure}{t}{0.45}
\subsection{The UV-to-Optical Colors at $z\sim4$.}

Our sample of $B-$dropouts has a mean redshift of $\langle
z\rangle=3.8\pm0.3$. We divide this sample based on their $i_{775}$
luminosity (which approximately corresponds to rest-frame 1500\,\AA)
in four bins of $\Delta i_{775} = 1$ mag.  These are the same bins as
shown in the histogram in Figure \ref{fig:magDist}.  The bins at
$i_{775}>28$ mag are too faint to be detected in $Spitzer$/IRAC and
will not be included in the analysis.  Assuming the mean redshift of
the sample, these bins correspond approximately to $M_{1500}=-21, -20,
-19,$ and $-18$, which covers the moderately bright to faint end of
the population ($M^*_{1600, z=4}=-20.24$, \citealt{bouw07}). 

In a recent study, \citet{lee11} focus on the median SEDs of the
brightest $z\sim4$ galaxies found in the ground based NOAO Deep
Wide-Field Survey. Their sample has a mean redshift $\langle z
\rangle=3.7\pm0.4$, and their SEDs also sample the rest-frame UV and
optical using a combination of optical filters: $B_W,\ R,\ I$ (Mosaic
Camera), near-IR filters: $J,\ H,\ K_s$ (NEWFIRM Camera), and the
$Spitzer$/IRAC mid-IR channels 1 through 4\footnote{Although
relatively deep IRAC channels 3 \& 4 data (sampling 5.8 $\mu$m and 8.0
$\mu$m respectively) is available for our sample in the GOODS-S, this
data is still $\gtrsim0.5\,$mag shallower than the channel 1 \& 2. It
is, therefore, much harder to obtain reliable stacked photometry in
those bands, especially considering that our sample is much smaller
and intrinsically fainter than the \citet{lee11} sample.  We have not
included these bands in the median-stacked SEDs presented in this
work.}. Similar to the present work, they divide their sample
according to the $I$--band luminosity, although with uneven bins.
Their sources are intrinsically brighter than in this work,
corresponding to $L>L^*_{z=4}$ sources roughly covering the range
$-23\lesssim M_{1500} \lesssim -21$.

The combination of the \citet{lee11} sample with our data spans the
UV-luminosity range: $-23\lesssim M_{1500} \lesssim -17.5$.  This
represents an unprecedented dataset to study the SEDs, and especially
the UV-to-optical colors, of $z\sim4$ galaxies over a large range of
luminosities.  The SEDs from both works are presented in Figure
\ref{fig:LeeComparisonSEDs}. In this figure the $H_{160}$ is shown
with open symbols since it is likely biased and should be ignored as
is explained in detail later. An overall trend to bluer colors (both
in the rest-frame UV and the rest-frame UV-to-optical colors) towards
fainter luminosities is already apparent in this figure.

We quantify the UV-to-Optical colors by estimating the interpolated
rest-frame $(U_n-V_n)$ colors for the median-stacked SEDs in both sets.
Figure \ref{fig:LeebetaandUtoV} shows a not very steep but systematic
trend of redder $(U_n-V_n)_{rest}$ colors as a function of increasing
UV-luminosity.

The interpolated $(U_n-V_n)$ colors were determined from the best fits
to the full observed SED, where the fits correspond to a linear
combination of a set of template SEDs. These SEDs correspond to the
default template set from the photometric code EAzY \citep{bram08}.
This template set was constructed and calibrated empirically to
determine accurate photometric redshifts of galaxies of various types,
from very blue starbursts to red ellipticals and has been shown to
work well up to $z\sim5-6$ \citep{bram08}.  For the $U_n$ and $V_n$
filters we have assumed narrow ideal filters that correspond to step
functions of width 100\,\AA, centered at 3500\,\AA\ and 5500\,\AA\ for
the $U_n$ and $V_n$ filters respectively. The colors were calculated
from the best fit combination of templates using the filters described
above.  To estimate the uncertainty in the $(U_n-V_n)_{rest}$
estimates, we perturbed  the photometry and the redshift of each
median-stacked SED within the allowed uncertainty, and a new fit was
obtained.  A new set of colors was measured from this best fit and the
procedure was repeated a large number of times to derive the
confidence intervals.  A similar procedure was used to derive
rest-frame $M_{1500}$ magnitudes for an ideal filter centered at
1500\,\AA\ and with a width of 100\,\AA.

For the data-point at $M_{1500}\sim-19$, only an upper limit can be
obtained due to the limitations of the template set used here. The
EAzY default template set only includes SEDs with
$(U_n-V_n)_{rest}\gtrsim0.3$. The SED at $M_{1500}\sim-19$ is likely
to have a $(U_n-V_n)_{rest}$ color of $\sim0.3\,$mag, so no reliable
lower limit can be obtained.  We choose to show a $2\,\sigma$ upper
limit. Other more sophisticated algorithms to determine interpolated
rest-frame colors for an observed SED yield very similar results
(e.g., the InterRest code, \citealt{tayl09}, which implements the
algorithm described in \citealt{rudn03}).  However, such algorithms
also depend on the template set resulting in similar limitations.

It is important to note that, when deriving this color from best fits,
we have ignored the $H_{160}-$band in our stacks and the $H$ and
$K_{s}-$bands in the \citet{lee11} stacks.  Given the redshift
distribution of the sample it is expected that for some fraction of
the sources (the ones at the lowest redshifts) the $H-$band fluxes
measured will include some rest-frame optical light.  As a result, the
estimated median rest-frame UV flux, will be biased towards larger
values.  To attempt a correction for this would require us knowing the
exact redshift distribution of the sample, but even then a correction
would be highly uncertain, so we opt to remove these photometric
points from all fits.

In recent works, \citet{bouw09, bouw11a} accurately determine the
UV-slope of LBGs at $z>3$ as a function of UV-luminosity and redshift.
For a star-forming stellar population, the UV-slope, parametrized by
$\beta$ ($f_\lambda \propto \lambda^\beta$), is most sensitive to the
dust content of the galaxy.  In consequence, the authors argue that
the dependence of $\beta$ on UV-luminosity can be fully explained by
an increase in the amount of dust that these galaxies contain as a
they become brighter.  If this is the case, a trend of redder
$(U_n-V_n)_{rest}$ colors for UV-brighter sources like the one observed
here is also expected, since this color is also affected by dust
extinction.  The bottom panel of Figure \ref{fig:LeebetaandUtoV} shows
the $(U_n-V_n)_{rest}$ colors corrected for dust extinction.  These were
derived as in \citet{bouw09}, from the mean $\beta$ vs. UV-luminosity
trend and the classic \citet{meur99} relation that relates the dust
content with the UV-slope $\beta$ according to:

$$A_{1600} = 4.43 - 1.99 \beta$$

where $A_{1600}$ is the extinction in magnitudes at 1600\,\AA.  A
\citet{calz00} extinction curve is used to estimate the extinction at
other wavelengths.

\colorsLee{figure}{h!}{0.45}
\allSEDs{figure*}{ht!}{0.95}
\tablePhotometry
\combSEDs{figure}{h}{0.45}
\figAgeDustModels{figure*}{!ht}{0.9}
\colorMag{figure}{h!}{0.45}

As can be seen from the figure, the dust corrected $(U_n-V_n)$ color
is essentially constant with UV-luminosity. This again suggests that
the main origin for the correlation of $\beta$ with the UV-luminosity
is an increasing dust content at brighter magnitudes.  Otherwise we
would expect some residual trend in the dust corrected color. This
would be the case for example, if a strong age trend existed with
UV-luminosity.  As will be shown later, a similar flattening of the
corrected colors is observed at all redshifts in the
$-21<M_{1500}<-18$ range.

Although both are flat, the trends of the corrected colors with
UV-luminosity of the two samples are disjoint.  The reasons for this
are unclear but there are a couple of reasons that may help explain
it.  First, it is unclear that the mean trend used to derive dust
corrections applies over such a large range of luminosities.  This can
help to explain that the corrected colors of the brightest sources are
bluer than those of the faintest ones.  However, it does not explain
the break between the two samples.  Second there are differences
between the selection of the two samples, in particular, it is not
clear that the fraction of contaminants is comparable between the two
samples. A larger fraction of contaminants in the brighter
ground-based sample of \citet{lee11} as compared to the our data,
could bias the brighter SEDs to smaller $(U_n-V_n)$ colors, resulting
in excessive corrections for dust, which are derived from the mean
trend on the UV with UV-luminosity. It should also be noted that in
the case of the \citet{lee11} sample, the UV-slope is not very well
sampled, with basically the $I$ and $J-$bands driving the UV-slope
(since the $H-$band has been excluded).  Even small systematic offsets
of $\sim0.1\,$mag in the zero-point calibration between these two
filters can bias the best fit models to redder slopes and in
consequence, smaller UV-to-optical colors.

\subsection{The UV-to-optical Colors at Higher Redshift}

\tableColors

We now extend our determination of the SED of high-$z$ galaxies to
$z\sim5$ and  $z\sim6$, and combine our determinations with the
stacked SEDs of sources at $z\sim7$ recently presented in
\citet{labb10}. Figure \ref{fig:SEDs} shows the SEDs at $4\lesssim z
\lesssim7$.  The expected mean redshifts of these samples correspond
to $\langle z\rangle=3.8, \langle z\rangle=5.0, \langle z\rangle=5.9,$
and $\langle z \rangle=6.9$, with typical spread in redshift of
$\Delta z \pm 0.3$. The samples at each redshift have been split in
bins of 1\,mag, using as reference the observed magnitude in the band
closest to 1500\,\AA\ (Table \ref{tbl:SEDS} contains the SEDs and
information about the samples in each bin).  Best fit BC03 models with
CSF and $0.2Z_\odot$ metallicity are overlaid for reference.  As
explained in the previous section, we have excluded from the fitting
the bands that are potentially biased (marked by the open symbols).
As can be seen in this Figure, all these SEDs are remarkably similar
and show sizable UV-to-optical colors.

The similarity of the SEDs can also be appreciated in Figure
\ref{fig:combSEDs}, where the SEDs from the different redshift samples
have been grouped according to their approximate $M_{1500}$
luminosities. The SEDs in the three $M_{1500}$ bins have been
re-normalized according to their $M_{1500}$ luminosity and combined to
produce over-sampled SEDs, which take advantage of the fact that the
different filters probe different rest-frame wavelengths at the
different redshifts.  The combined SEDs show the typical flat UV
slopes, but also rather flat colors in the rest-frame optical, albeit
with a relatively large scatter.  This scatter can in part be a
consequence of the normalization which is done in the far UV, which
effectively acts as the pivot point. Different emission lines that
sometimes contribute flux in the IRAC bands, depending on the
redshift, can also be contributing to this scatter.

\subsubsection{Emission lines in the rest-frame optical.}

We first consider the implications if the sizable red $J_{125}-[3.6]$
colors that we observe are assumed to be exclusively measuring the
stellar continuum in these high redshift galaxies. The red color would
arise from large Balmer breaks, and would be indicative of evolved
stellar populations. As can be seen in Figure \ref{fig:ageModels}a,
the $J_{125}-[3.6]$ colors characteristic of these sources would then
require fairly large ages of $\sim1$ Gyr if no dust is allowed in the
models (the model tracks in the Figure are CSF models).  Allowing some
reddening by dust (using in this case a \citealt{calz00} law), reduces
their ages considerably (Figure \ref{fig:ageModels}c). The amount of
dust reddening is constrained however.  Based on the UV continuum
slopes typically observed in these galaxies \citep{bouw09}, and using
the $z=0$ \citet{meur99} relation between this slope and the dust
content, it has been shown that the reddening at $z\gtrsim4$ is
typically $A_V\lesssim 0.7\,$mag. The reddening is likely to be even
smaller at higher redshifts \citep{bouw09, bouw11a}.  With this added
constraint, the ages derived from SED fitting are generally estimated
to be $300-400$ Myr \citep[e.g.,][]{yan05, eyle05, star09, gonz10,
labb10}.

In our SEDs at $z\gtrsim5$ we also find that the $([3.6]-[4.5])$ color
is consistently blue, with typical values of $\sim-0.3\,$mags.  This
color is very hard to reproduce with a model with any star formation
history that is based only on stellar continuum.  The addition of dust
only makes the comparison worse (Figure \ref{fig:ageModels}c,d). It
should be noted, however, that the uncertainties in this color are
considerable and that in most cases the observed color is formally
consistent with the stellar continuum only models.

It has been claimed \citep{scha09, scha10} that at some of these
redshifts, particularly $z\sim6$ and 7, the SEDs could be better fit
by very young stellar population models ($\lesssim10$ Myr old) with
moderate dust content and very strong nebular emission lines.  These
lines would dominate the rest-frame optical fluxes. As the galaxies
are forming stars, it is likely that they have nebular emission lines,
and so we have assessed whether emission lines could be influencing
our fluxes and fits. The observed position of the most prominent
optical nebular emission lines at each redshift is marked in Figure
\ref{fig:SEDs} by the dashed vertical lines (assuming the mean
redshift of each sample).  It can be seen that at the different
redshifts involved, some of the rest-frame optical emission lines
would lie within the IRAC [3.6] and [4.5] filters.

\citet{atek11} showed recently that at $z<2.8$ there are sources that
show emission lines with large enough equivalent widths to dominate
the broadband fluxes in the optical.  However, even for their sample
of high equivalent width selected sources, such objects are not very
common. For the more typical cases, the lines that they find in their
sample would have moderate contributions to the IRAC broad bands,
usually making them $0.2-0.3\,$mags brighter. At higher redshifts,
$3.8\lesssim z\lesssim5$, the work of \citet{shim11}, based on
spectroscopic redshifts and IRAC broadband photometry, finds similar
results.  They find that the sources in this redshift range show an
excess of flux in [3.6] over [4.5], which the authors think can be
explained by H$\alpha$ emission. The majority of the sources in their
sample have $([3.6]-[4.5])$ colors that are consistent with an
H$\alpha$ contribution of $0.2-0.3\,$mags to the [3.6] filter,
indicating moderate contributions from emission lines.

We consider the effects of a similar (moderate) contribution of
emission lines to the colors of a 280\,Myr old CSF model. For the
purpose of this exercise we assume a H$\alpha$ rest-frame equivalent
width ${\rm EW_{rest}}=300\,$\AA. At $z\sim6$, this would increase the
luminosity in the [4.5] filter by 0.23\,mag. We only consider the
strongest lines in the optical and assume the strength ratios
presented by \citet{ande03} for a $0.2\,{\rm Z_\odot}$ metallicity
system. This corresponds to ${\rm EW_{rest}(OII, H\beta, OIII) = (189,
105, 670)\,}$\AA, respectively. The open squares in Figure
\ref{fig:ageModels}a,b, correspond to the median colors for a model
like the one described (the median takes into account the expected
redshift distribution of each sample).

As can be seen in Figure \ref{fig:ageModels}a, the main effect of the
contribution from the emission lines is that it allows for
significantly younger ages to reproduce the $J_{125}-[3.6]$ colors (by
roughly a factor of two). Since this implies lower stellar continuum
fluxes than previously assumed, this would also result in lower best
fit stellar masses than previously estimated for galaxies at these
redshifts (by a similar factor).  Simultaneously, given the line
ratios and the redshift distributions expected for these samples, a
model like this can also help in reproducing the blue [3.6]-[4.5]
colors of the SEDs at $z\gtrsim5$ (Figure \ref{fig:ageModels}b). While
formally models with just stellar continua can fit the data, given the
current large uncertainties, such colors are hard to obtain with
stellar continuum only models. It is more likely that some degree of
contribution of emission lines provides a better solution.

The assessment presented here is necessarily very simplistic, given
the current uncertainties, and is only meant to show the possible
effects of emission lines on the colors of these galaxies.  Better
estimates of the contribution of these lines and its effects on the
properties derived from SED fitting will depend on the actual
strengths and ratios of the lines. Moderate improvement can be
achieved if more precise redshifts are known for these sources, such
data does not exist at this time for such faint high redshift galaxies
and so this remains an open question until improved spectroscopic
capability becomes available.

\subsubsection{UV-to-optical color vs. luminosity.}

Regarding the trend of redder UV-to-optical colors for brighter
UV-luminosities presented in Figure \ref{fig:LeebetaandUtoV} for the
$z\sim4$ sample, it can also be observed for the other samples.  In
fact, within the uncertainties in the color determination, the
$J_{125}-[3.6]$ vs $M_{1500}$ trend is roughly the same at $z\sim5$
and 6 (Figure \ref{fig:colorMag}, \emph{top}).  This is particularly
remarkable in view of the fact that this color is an observed color,
not a rest-frame color, and in consequence is probing different
wavelengths for the different samples.  This shows again how similar
and flat these SEDs are. If all the SEDs are considered
simultaneously, a trend of $J_{125} - [3.6] = -0.12(\pm0.07) \times
M_{1500,AB}+const$, is found. For simple CSF models, this trend
implies a variation in stellar M/L ratio.  \citet{gonz11} find that at
$z\sim4$, this variation is of a factor $\sim\times5$ between
$M_{1500}=-18$ and $M_{1500}=-21$. So, despite the steep UV-luminosity
functions characteristic at $z\gtrsim4$, the contribution of the
faintest sources to the total stellar mass density is more modest than
their contribution to the star formation rate density.

Similar to the analysis performed at $z\sim4$, we have derived
rest-frame $(U_n-V_n)$ colors for the median SEDs at all redshifts.
These colors are shown in Figure \ref{fig:colorMag}. The trend with
UV-luminosity has large scatter and low significance but is systematic
in the sense that brighter sources are redder in $(U_n-V_n)_{rest}$.
A fit to the rest-frame colors as a function of M$_{1500}$ of all
redshifts simultaneously, results in $(U_n-V_n) = -0.04\times M_{1500}
- 0.37$, shown as the brown dashed line in Figure \ref{fig:colorMag}
(\emph{center} panel). This is slightly steeper than the $-0.03$ slope
found for the $z\sim4$ data only, also plotted for reference as the
black dashed line in the same Figure. The latter trend included the
bright sample of \citet{lee11} and is consistent within the
uncertainties with the fit to all the points.

Using the median trends determined by \citet{bouw11a} regarding the
$\beta$ slopes vs. UV-luminosity for $4\lesssim z\lesssim7$ LBGs, we
corrected the  $(U-V)_{rest}$ colors for dust extinction using the
\citet{meur99} relation (Figure \ref{fig:colorMag}, \emph{bottom}).
These corrected colors show large scatter around $(U-V)_{rest,
corrected}\sim0.4$ but there is no indication of a residual trend with
UV-luminosity except for the $z\sim7$ sample (black symbols).  This
may indicate again that at $4\lesssim z\lesssim6$, the evolution can
be fully explained by a change in dust content and there is no strong
evolution in the ages of these sources as a function of UV-luminosity.
This is consistent with the report by \citep{star09} based on stellar
population modeling of slightly brighter LBGs in this redshift range.
It is also consistent with the predictions from the hydrodynamical
simulations of \citet{finl11}, which find no trend of age with
UV-luminosity.

At $z\sim7$ there seems to be a residual trend in  $(U_n-V_n)_{rest,
corrected}\sim0.4$ vs. UV-luminosity in the sense that brighter
sources have systematically bluer intrinsic $(U_n-V_n)$ colors.  In
principle, this could indicate that the brighter sources are actually
younger than the fainter sources, which would be difficult to imagine.
Alternatively, it is possible that the simple dust corrections
(calibrated at $z=0$) are not adequate at these high redshifts, due
to, for example, changes in the IMF, metallicity, or escape fraction,
all of which could alter the UV-slope $\beta$ for a given dust
content.  It should also be noted that the $\beta$ vs. UV-luminosity
at $z\sim7$ is considerably steeper than at the other redshifts and so
the corrections to the $(U_n-V_n)_{rest}$ vs. UV-luminosity trend are
more extreme. Nevertheless, the uncertainties are sufficiently large
that also the $z\sim7$ trend is fully consistent with being constant.

\section{Summary}

Determining the rest-frame UV-to-optical colors of individual UV-faint
galaxies at $z\gtrsim4$ is challenging.  At such redshifts the
rest-frame optical lies at $\lambda>2.5$ microns in the mid-IR and so
it is hard to measure at the extremely faint magnitudes typical of
galaxies in the first 1.5 Gyr. To gain insight into the typical
spectral properties of $z\sim4-6$ galaxies, we have taken advantage of
the large samples of $z\sim4-6$ sources found in the deep HST ACS and
WFC3/IR images of the ERS and HUDF fields. The HST data give us
rest-frame UV fluxes.  We then use $Spitzer$/IRAC [3.6] and [4.5]
micron data from the deep GOODS survey of these fields to get
rest-frame optical fluxes by stacking the IRAC images and determining
median flux values. This allows us to determine the SEDs of these
galaxies covering both the rest-frame UV and the optical.  These SEDs
represent a first comprehensive study of the rest-frame UV-to-optical
properties of $z\gtrsim4$ galaxies as a function of both redshift and
UV-luminosity down to very faint limits.

At $z\gtrsim4$ we have also combined our faint SEDs with the $L>L^*$
stacks presented in \citet{lee11}.  This allows us to assess the
colors of $z\gtrsim4$ sources over an unprecedentedly large range of
luminosities. Our main findings are as follow:

\begin{itemize}

  \item At $z\gtrsim4$ the $(U_n-V_n)$ rest-frame color (interpolated
    from the data using a fitted SED from a set of templates is of
    $(U_n-V_n) \sim 0.5\,$ mags. There is a shallow but systematic
    trend of redder colors for brighter UV-luminosities over the range
    $-23\lesssim M_{1500,AB} \lesssim -17.5$ (Figure
    \ref{fig:LeebetaandUtoV}, section 5.1). A linear fit to this trend
    is $(U_n-V_n)_{rest}=-0.03\times$M$_{1500}-0.07$

  \item The SEDs of star forming galaxies at $4\lesssim z \lesssim7$
    are remarkably similar at all luminosities from $-23\lesssim
    M_{1500,AB} \lesssim -17.5$, showing fairly flat rest-frame UV and
    rest-frame optical colors (section 5.2 and Figures \ref{fig:SEDs},
    \ref{fig:combSEDs} and \ref{fig:colorMag}). At a given redshift,
    there are weak indications of a subtle trend for redder colors for
    brighter sources. A simple fit to the data at all redshifts
    simultaneously results in
    $(U_n-V_n)_{rest}=-0.04\times$M$_{1500}-0.37$ (Figure
    \ref{fig:colorMag} \emph{center}, Section 4.2).

  \item The UV-to-optical color, as measured by the observed
    $J_{125}-[3.6]$, color remains fairly constant with redshift,
    despite the $Spitzer$/IRAC bands probing different wavelength
    regions of the SED (Figure \ref{fig:colorMag} \emph{top}). This
    suggests that the optical colors are fairly flat at all redshifts
    $z\sim4-7$, although there is significant scatter, which can be
    caused by optical emission line contamination to the IRAC filters.
    
  \item The $z\gtrsim5$ SEDs show consistently blue
    $[3.6]-[4.5]\sim-0.3$ colors. This is hard to reproduce with
    models that only include stellar continuum, although it is still
    formally possible given the current uncertainties.  Nonetheless,
    it is probably better explained with moderate flux contributions
    from optical nebular emission lines in the two IRAC bands.
    Including some contribution from emission lines leads to lower
    best fit ages and lower stellar masses than previously estimated
    for galaxies at these redshifts. Changes of a factor 2 could well
    result, but we would caution that detailed assessments are not yet
    possible with the current data.
  
  \item We derive dust-corrected $(U - V)$ colors using a simple dust
    correction based on the median UV-slope $\beta$ vs. UV-luminosity
    relation from \citet{bouw09, bouw11a} and the \citet{meur99}
    relation (Figure \ref{fig:colorMag}, \emph{bottom}). The
    dust-corrected UV-to-optical colors at $4\lesssim z\lesssim6$ are
    $(U-V)\sim0.4\,$mags. They do not appear to depend on
    UV-luminosity.  This may suggest that moderate changes in the dust
    content of galaxies can explain the color dependency observed both
    in the UV-slopes and the UV-to-optical colors. In particular, it
    seems that no age evolution is required to match this color
    dependency.

\end{itemize}

The recent deep and ultra-deep HST ACS and WFC3/IR imaging programs
over the GOODS-S field have resulted in large samples of $z\sim4$
galaxies. Taking advantage of these large samples, along with a
careful stacking analysis, we present a first comprehensive study of
the typical SEDs over the rest-frame UV and the rest-frame optical for
faint star forming galaxies in the $4\lesssim z \lesssim 7$ redshift
range. The stacked SEDs allow us to study the UV-to-optical colors of
high-z star forming galaxies down to very low luminosities.  We find a
weak trend to bluer color at fainter luminosities.  Interestingly,
these stacked SEDs also show a remarkable similarity across redshift
in a period of considerable stellar mass growth from about 0.8 Gyr to
1.5 Gyr, suggesting a smooth self-similar mode of evolution at $z\sim
7 - 4$.

\acknowledgements

We acknowledge support from NASA grant HST-GO-11563, and NASA grant
HST-GO-11144. This work is based on observations made with the Hubble
Space Telescope and with the Spitzer Space Telescope. Support for this
work was provided by NASA through an award issued by JPL/Caltech. P.O.
acknowledges support provided by NASA through Hubble Fellowship grant
HF-51278.01 awarded by the Space Telescope Science Institute, which is
operated by the Association of Universities for Research in Astronomy,
Inc., for NASA, under contract NAS 5- 26555.

\clearpage
\newpage

\begin{appendix}

\section{A. Depth of the Stacks}

To  quantify the rest-frame optical properties of very faint
high-redshift galaxies, we stack the IRAC images for a large number of
faint galaxies, after removing the flux contribution from nearby
neighbors (section 4).  While we do not expect large systematics in
the photometry of individual sources from our deblending procedure to
remove neighboring sources, it is possible that modest systematics
could arise when stacking a large number of sources.  The role of this
appendix is to determine how large such systematics might be.
Fortunately we are able to demonstrate that the fluxes that we measure
in the stacked IRAC images are unbiased and that their significance is
properly determined.

\begin{figure}[!h] \centering
  \includegraphics[width=0.8\textwidth]{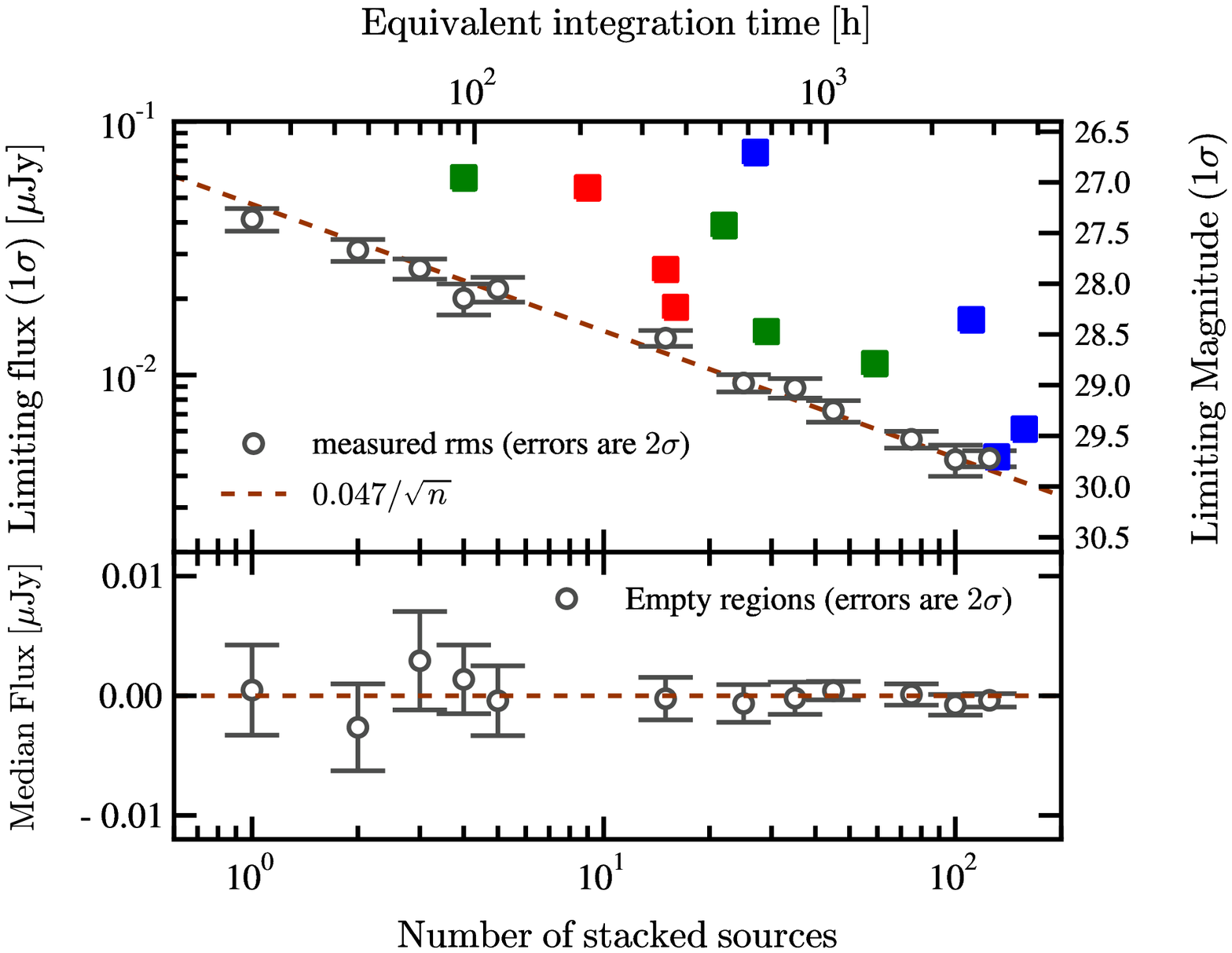}
  \caption{\emph{Upper:} The $1\,\sigma$ limiting flux (in $\mu$Jy) as
  a function of the number of IRAC stamps stacked.  All the stamps
  have the same integration time of $\sim23.4\,$h. The estimates (open
  symbols) come from stacking large numbers of empty areas of the
  GOODS images.  Each empty area was cleaned of neighboring fluxes
  using the code described in the text.  The noise decreases $\propto
  \sqrt{n}$, with n being the number of empty regions stacked.  The
  filled symbols are the actual RMS determined for each of our stacked
  stamps: blue for the $z\sim4$ stamps, green $z\sim5$, and red
  $z\sim6$ (see Figures \ref{fig:BstackStamp}, \ref{fig:VstackStamp},
  and \ref{fig:IstackStamp}).  These RMS values are $\gtrsim$ than the
  ones derived for empty regions. This is expected because the
  uncertainties in the stack include both the image noise and
  intrinsic variations in the population being stacked. \emph{Lower:}
  The flux of the median combined stacks of cleaned empty regions as a
  function of the number of stamps combined. As expected, this is very
  close to zero.} \label{fig:stackLimits} \end{figure}

We started by selecting 200 empty areas in the HUDF field.  We select
them based on the segmentation maps derived from the high resolution
HST images.  We processed each of the selected empty areas as if they
were the position of one of the real sources in our catalog, i.e. we
ran them through the de-blending code to remove the flux from nearby
un-associated sources.  To determine the rms for a given $N_{stacked}$
number of stacked stamps, we randomly draw (with replacement)
$N_{stacked}$ stamps from the 200 empty areas, median combine them and
perform aperture photometry in the median stack in the same fashion as
for our real stacks.  We repeat the drawing 300 times and measure the
rms of these 300 trials. As can be seen in the upper panel of Figure
\ref{fig:stackLimits}, the limiting flux decreases as $\propto
1/\sqrt{N_{stacked}}$, as expected for background-limited noise.
In the case of the stacks that contain real sources, the rms is larger
than the image noise because it is also affected by the intrinsic
variation in the distribution of fluxes that are being stacked.

As can be seen in the lower panel, the flux of the median stacks is
very close to the expected zero flux.

\end{appendix}

\end{document}